%% file: sample-manuscript.tex
\documentclass[manuscript,screen,final]{acmart}
\AtBeginDocument{%
  \providecommand\BibTeX{{%
    \normalfont B\kern-0.5em{\scshape i\kern-0.25em b}\kern-0.8em\TeX}}}

\setcopyright{acmlicensed}
\copyrightyear{2025}
\acmYear{2025}
\acmDOI{XXXXXXX.XXXXXXX}
\acmJournal{TOIS}
\acmVolume{XX}
\acmNumber{X}
\acmArticle{XXX}
\acmISBN{978-1-4503-XXXX-X/18/06}

\usepackage{subfig}
\usepackage{algorithm}
\usepackage{algorithmic}
\usepackage{makecell}   % 三线表-竖线
\usepackage{booktabs}   % 表格
\usepackage{tabularx}   % 表格
\usepackage[table]{xcolor}
\usepackage{multirow} 
\begin{document}

\title{Why not Collaborative Filtering in Dual View? Bridging Sparse and Dense Models}

\author{Hanze Guo}
\affiliation{%
  \institution{Gaoling School of Artificial Intelligence, Renmin University of China}
  \city{Beijing}
  \country{China}
    \postcode{100872}
    }
\email{ghz@ruc.edu.cn}
  
\author{Jianxun Lian}
\affiliation{%
  \institution{Microsoft Research Asia}
  \city{Beijing}
  \country{China}
    \postcode{100872}
}
\email{jianxun.lian@outlook.com}

\author{Xiao Zhou}
\authornote{Corresponding author}
\affiliation{%
  \institution{Gaoling School of Artificial Intelligence, Renmin University of China}
  \city{Beijing}\country{China}}
\additionalaffiliation{%
  \institution{Beijing Key Laboratory of Research on Large Models and Intelligent Governance}
  \city{Beijing}\country{China}}
\additionalaffiliation{%
  \institution{Engineering Research Center of Next-Generation Intelligent Search and Recommendation, MOE}
  \city{Beijing}\country{China}}
\email{xiaozhou@ruc.edu.cn}

\renewcommand{\shortauthors}{Guo, et al.}

\begin{abstract}
Collaborative Filtering (CF) remains the cornerstone of modern recommender systems, with dense embedding--based methods dominating current practice. However, these approaches suffer from a critical limitation: our theoretical analysis reveals a fundamental \emph{signal-to-noise ratio (SNR) ceiling} when modeling unpopular items, where parameter-based dense models experience diminishing SNR under severe data sparsity. To overcome this bottleneck, we propose \textbf{SaD (Sparse and Dense)}, a unified framework that integrates the semantic expressiveness of dense embeddings with the structural reliability of sparse interaction patterns. We theoretically show that aligning these dual views yields a strictly superior global SNR. Concretely, SaD introduces a lightweight bidirectional alignment mechanism: the dense view enriches the sparse view by injecting semantic correlations, while the sparse view regularizes the dense model through explicit structural signals. Extensive experiments demonstrate that, under this dual-view alignment, even a simple matrix factorization--style dense model can achieve state-of-the-art performance. Moreover, SaD is plug-and-play and can be seamlessly applied to a wide range of existing recommender models, highlighting the enduring power of collaborative filtering when leveraged from dual perspectives. Further evaluations on real-world benchmarks show that SaD consistently outperforms strong baselines, ranking first on the BarsMatch leaderboard.\footnote{\url{https://openbenchmark.github.io/BARS/Matching/leaderboard/index.html}} The code is publicly available at \url{https://github.com/harris26-G/SaD}.

\end{abstract}





\maketitle

\section{Introduction}

% Recommender systems are widely deployed in e-commerce~\cite{linden2003amazon,zhou2018deep}, video streaming~\cite{covington2016deep}, and news services. They effectively manage information overload by delivering personalized content based on user preferences. By analyzing user behavior to predict relevant items, these systems significantly enhance user satisfaction and engagement. As a core technique, Collaborative Filtering (CF) leverages historical interactions to infer preferences, fundamentally shaping item discovery across domains.

Recommender systems are widely deployed in e-commerce~\cite{linden2003amazon,zhou2018deep}, video streaming~\cite{covington2016deep}, and news services. They effectively manage information overload by delivering personalized content based on user preferences. By analyzing large-scale user behavior, these systems uncover patterns to predict relevant items, enhancing user satisfaction and key business metrics like engagement and retention. As a core technique, Collaborative Filtering (CF) leverages historical interactions to infer preferences, fundamentally shaping item discovery across domains.

\begin{figure}[htbp!]
  \centering
  \subfloat[Performance on Yelp]{%
    \includegraphics[width=0.48\linewidth]{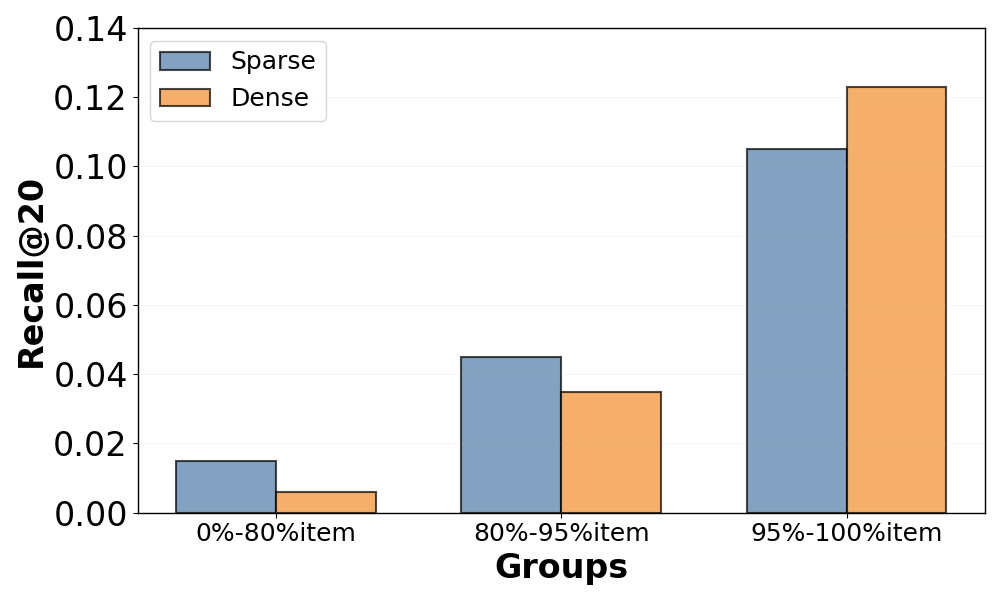}%
    \label{fig:image1}%
  }\hfill
  \subfloat[Performance on Movielens]{%
    \includegraphics[width=0.48\linewidth]{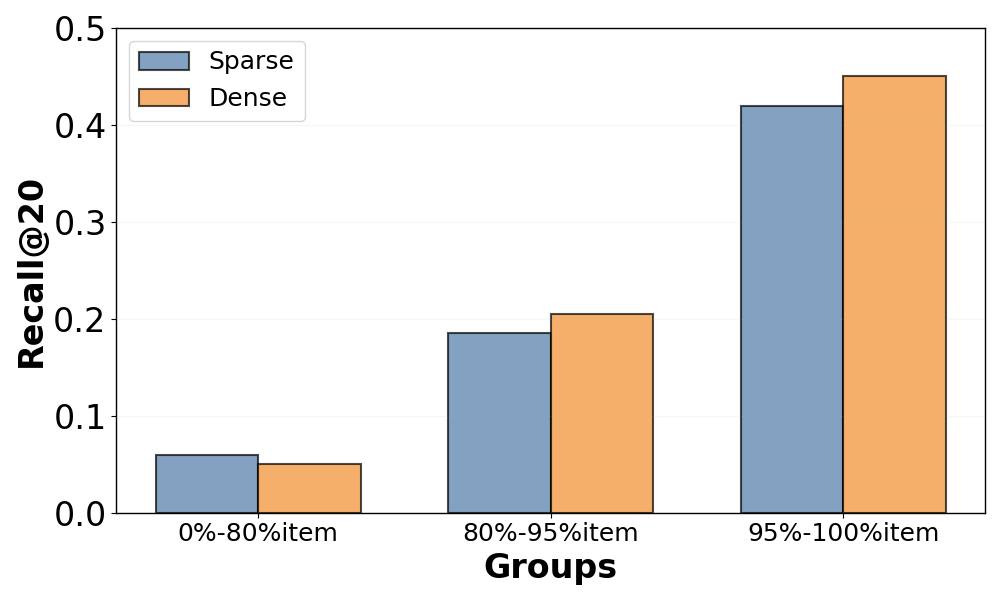}%
    \label{fig:image2}%
  }
  \caption{Items are grouped according to the number of user interactions and divided into three categories, from the least to the most interactions. Owing to the low signal-to-noise ratio (SNR), commonly used dense models perform poorly on unpopular items.}
  \label{fig:0}
\end{figure}

Early CF models, such as matrix factorization, learn dense embeddings for users and items~\cite{cheng2016wide}. Recently, Graph Neural Networks (GNNs)~\cite{ying2018graph,berg2017graph,zhou2025tricolore,guo2025sorex,ma2024tail,li2025leave} have gained prominence for capturing high-order interactions, often outperforming their predecessors~\cite{chen2020revisiting,ji2020dual,sun2020neighbor,zhou2019topic}. Despite this progress, data sparsity remains a critical bottleneck, as embedding-based methods struggle to learn robust representations from limited interactions. To mitigate this, self-supervised learning (SSL) techniques have been introduced to augment training signals via generated data or auxiliary objectives. By exploring diverse information domains~\cite{he2020momentum,guo2025counterfactual}, SSL enhances embedding quality and model generalization.

% Early CF models employ matrix-factorization or two-tower architectures that learn dense embeddings for users and items~\cite{cheng2016wide}. More recently, models based on Graph Neural Networks (GNNs)~\cite{ying2018graph,berg2017graph} have gained attention for their ability to capture higher-order user-item interactions using graph structures~\cite{huang2021mixgcf}, often outperforming earlier models~\cite{chen2020revisiting,ji2020dual,sun2020neighbor}. Despite this progress, the sparsity of user-item pairs in recommender systems poses a significant challenge. Embedding-based methods often struggle to capture user and item features effectively due to limited interactions. To address this, self-supervised learning techniques enhance learning by generating additional data or using diverse loss functions, helping models better understand user and item characteristics. Self-supervised loss broadens the model's learning space, enabling it to find better solutions. Methods like those in~\cite{he2020momentum} shift the model into a different information domain, improving embedding quality. 

While self-supervised learning mitigates sparsity~\cite{he2020momentum}, it operates primarily within the latent embedding view, overlooking the potential of explicit structural priors. Our analysis reveals that embedding-based methods and structure-based models exhibit different characteristics. Embedding-based methods inherently suffer from a low Signal-to-Noise Ratio (SNR) on sparse interactions, whereas structure-based models leverage local connectivity to maintain robustness. As illustrated in Figure~\ref{fig:0}, the dense embedding-based model (MF) tends to underperform on less popular items, whereas traditional structure-based methods (slim) achieve more robust results in these scenarios. A similar trend is further confirmed in our plug-and-play analysis in Section~\ref{sec:generalizability}, where applying the proposed SaD strategy consistently improves performance across backbones, particularly for long-tail items.

Motivated by this observation, we categorize collaborative filtering models into two complementary views: the \textbf{Sparse View} and the \textbf{Dense View}. \textbf{Sparse-View methods} (e.g., ItemCF~\cite{linden2003amazon}, slim~\cite{ning2011slim}) operate directly on the explicit interaction graph without learning latent embeddings. While they are robust under data sparsity, they lack the capacity to capture higher-order semantics. In contrast, \textbf{dense-view methods}, such as LightGCN~\cite{he2020lightgcn}, learn expressive user and item embeddings and excel at modeling complex relational signals. However, their reliance on sufficient interaction data renders them less effective when facing long-tail items. Moreover, dense models often struggle to represent simple co-occurrence patterns, which are naturally encoded in the sparse view.

Integrating these views remains non-trivial due to the \textbf{semantic gap} between continuous latent embeddings and discrete interaction graphs. While recent works~\cite{mao2021ultragcn,shen2021powerful,liu2023personalized} attempt to combine hybrid components, they \textbf{fail to recognize the fundamental differences} between sparse and dense models, preventing principled alignment between the two. More importantly, these methods lack a \textbf{theoretical understanding} of how fusion reshapes the optimization landscape (e.g., through signal-to-noise ratio improvement), resulting in suboptimal and unprincipled designs. Furthermore, most existing approaches are highly \textbf{model-specific}, limiting their transferability across diverse recommendation frameworks. These limitations highlight the need for a unified and theoretically grounded dual-view approach. These limitations underscore the need for a unified approach. To our knowledge, this work is the first to provide a principled framework that explicitly aligns these dual structures for collaborative filtering.

%To address this issue, we first demonstrate theoretically that incorporating sparse view models effectively improves the signal-to-noise ratio (SNR) in dense view predictions for inherently sparse user-item (U-I) matrices. We then propose integrating sparse view methods into dense view embedding-based approaches. These sparse view methods do not rely on explicit learning processes, which enhances the effectiveness of the embedding-based techniques.
%By combining the strengths of both dense view and sparse view methods, we improve the model’s ability to capture information at multiple levels. Finally, we introduce a dual-view model that integrates both structural information and embedding features, further enhancing performance.

To address this issue, we first provide a theoretical analysis showing that incorporating sparse-view models can effectively improve the signal-to-noise ratio (SNR) of dense-view predictions for inherently sparse user–item (U–I) interaction matrices. Motivated by this result, we propose integrating sparse-view methods into dense embedding-based approaches. Unlike dense models, sparse-view methods do not rely on learning dense latent representations, which enables them to provide stable structural signals that complement embedding-based techniques. By combining the strengths of both views, the resulting model captures information at multiple structural and semantic levels.

%Specifically, we propose the \textbf{Sparse and Dense (SaD)} framework, which comprehensively captures recommendation system information through the alignment of two different views. The model integrates both the sparse structure to capture simple user-item relationships and the dense structure to capture user and item characteristics. By capturing information from two different views, the model achieves better performance. We also design a model interaction mechanism that facilitates learning between the different views. The dense view aids the sparse view, while the sparse view helps the dense model. Our approach employs a label-augmentation method to enhance the learning process of the dense model and refines the sparse model by incorporating dense embedding information. Under this framework, even using simple sparse and dense models can lead to exceptional performance.

Specifically, we propose the \textbf{Sparse and Dense (SaD)} framework, which unifies collaborative filtering through principled alignment of sparse and dense views. The sparse view captures explicit co-occurrence and local structural patterns in user–item interactions, while the dense view models user and item characteristics via expressive embeddings. To bridge the semantic gap between these views, we design a bidirectional interaction mechanism that enables mutual enhancement: the dense view augments the sparse view with semantic signals, while the sparse view regularizes the dense model with explicit structural supervision. In particular, we introduce a label-augmentation strategy to improve dense model training and refine sparse modeling using dense embedding information. Under this framework, even simple sparse and dense backbones can achieve strong performance.

%Extensive experiments on four widely used datasets demonstrate the favorable results of our model. Our model also shows significant improvements in recommending less popular items, which is typically a challenge for many recommender systems. Meanwhile, our performance gains are not only substantial but also orthogonal to the advances in self-supervised learning, further highlighting the plug-and-play nature of our framework. Moreover, we validated the model’s generalization ability by evaluating it on additional datasets, confirming its robustness across diverse data.

Extensive experiments on four widely used datasets demonstrate the effectiveness of SaD. Our framework consistently improves recommendation accuracy, particularly for long-tail items, which are notoriously challenging for existing methods. Moreover, the observed performance gains are orthogonal to advances in self-supervised learning, highlighting the plug-and-play nature of SaD. Additional experiments on unseen datasets further confirm the robustness and generalization ability of the proposed framework.

%The main contributions of the paper are as follows: 
The main contributions of this paper are summarized as follows:

\begin{itemize}

%\item We identify and analyze the performance divergence between Sparse and Dense models. Through theoretical analysis, we demonstrate the superiority of integrating dual views. Based on this observation, we propose the idea of utilizing dual views to design a more effective model.
\item We identify and theoretically analyze the performance divergence between sparse and dense collaborative filtering models, and show that principled dual-view integration leads to superior optimization properties.

%\item We present a plug-and-play framework named SaD, which effectively integrates Sparse and Dense models and facilitates cross-view alignment between the dual views to enhance collaborative filtering performance, offering a new perspective to improve recommendation model performances.
\item We propose SaD, a plug-and-play dual-view framework that enables effective cross-view alignment between sparse and dense models, offering a unified perspective for improving collaborative filtering.

%\item Our model achieves state-of-the-art performance on four datasets, surpassing all models on the public datasets of the BARS leaderboard. The model also significantly improves the recommendation performance of unpopular items.
\item Extensive experiments demonstrate state-of-the-art performance on four benchmarks, including top results on the public BARS leaderboard, with particularly strong gains on unpopular items.

\end{itemize}

\section{Related Work}
\input{Ours/chapter/related_work}

\section{Preliminaries}
\input{Ours/chapter/preliminary}

\section{Methodology}\label{method}
\input{Ours/chapter/methods}

\section{Experiments}\label{experiment}

\input{Ours/chapter/experiments}

% \section{Result and Discussion}
% \input{chapter/results}

\section{Conclusion}

\input{Ours/chapter/conclusion}

\bibliographystyle{ACM-Reference-Format}
\bibliography{sample-base.bib}

% \appendix









\end{document}

%% file: Ours/chapter/related_work.tex
Collaborative filtering is a foundational approach in recommender systems, widely used for predicting user preferences based on historical interactions. In this study, we focus on ID-based collaborative filtering, where neither side information (e.g., user/item features) nor temporal context is available.

We categorize existing models into two types: \textbf{dense models} and \textbf{sparse models}. Dense models typically learn latent representations for users and items via matrix factorization~\cite{koren2009matrix} or deep learning methods, constructing a dense user-item interaction matrix for preference prediction. In contrast, sparse models directly exploit user-item co-occurrence patterns to construct item-item similarity matrices, which are used to infer user preferences by identifying items similar to those a user has previously interacted with.

\subsection{Dense Models for Collaborative Filtering}
%We begin by reviewing representative dense models. Matrix Factorization (MF)~\cite{koren2009matrix}, a fundamental embedding-based model, learns parameters through hidden vectors for result prediction. ENMF~\cite{chen2020efficient} (Explicit Neural Matrix Factorization) is a model framework that can learn model parameters from training data without the need for sampling, and it demonstrates excellent performance on the data. NCF (Neural Collaborative Filtering)~\cite{he2017neural} is a deep learning-based recommendation model that utilizes neural networks to capture user-item interactions and make personalized recommendations.

We begin by reviewing representative dense collaborative filtering models. Matrix Factorization (MF)~\cite{koren2009matrix}, a foundational embedding-based approach, learns latent user and item representations to predict user–item interactions. ENMF~\cite{chen2020efficient} (Explicit Neural Matrix Factorization) is an efficient framework that directly optimizes model parameters from training data without negative sampling, and has demonstrated strong empirical performance. Neural Collaborative Filtering (NCF)~\cite{he2017neural} extends MF by leveraging neural networks to model nonlinear user–item interaction patterns, enabling more expressive personalized recommendations.

%NGCF~\cite{fan2019graph} first introduces graph neural networks into the realm of recommender systems, achieving superior performance compared to traditional deep learning methods. Graph neural network methods like LightGCN~\cite{he2020lightgcn} simplify the structure of graph neural networks, enhancing model performance by learning neighbor information while reducing model complexity. SGL~\cite{wu2021self} improves model performance by integrating self-supervised learning into the framework of graph neural networks, while also enhancing the model's performance on niche products. SimpleX~\cite{mao2021simplex} demonstrates the importance of sampling strategies in effective learning and highlights the significant impact of negative sampling quantity on model performance. HCCF~\cite{xia2022hypergraph} utilizes hypergraphs to model collaborative filtering tasks. Singular Value Decomposition (SVD)~\cite{ma2016diffusion,narang2013signal} is widely used by researchers due to its effectiveness and adaptability across various fields. Additionally, there is work being done using hypergraphs~\cite{xia2022hypergraph} to model user-item relationships.

NGCF~\cite{fan2019graph} is among the first to introduce graph neural networks (GNNs) into recommender systems, achieving superior performance over traditional deep learning approaches. Subsequent GNN-based methods, such as LightGCN~\cite{he2020lightgcn}, simplify the network architecture to focus on neighborhood aggregation, improving effectiveness while reducing model complexity. SGL~\cite{wu2021self} further enhances GNN-based recommendation by incorporating self-supervised learning, leading to notable improvements on long-tail items. SimpleX~\cite{mao2021simplex} highlights the critical role of sampling strategies in effective training, demonstrating the significant impact of negative sampling on model performance. Beyond standard graphs, HCCF~\cite{xia2022hypergraph} leverages hypergraph structures to model higher-order relationships in collaborative filtering. In addition, Singular Value Decomposition (SVD)~\cite{ma2016diffusion,narang2013signal} remains a widely adopted technique due to its effectiveness and adaptability across diverse domains.

%As GNN-based models often suffer from over-smoothing — making user and item embeddings indistinguishably close — recent methods adopt self-supervised and contrastive learning to reshape the embedding distribution. Building upon SGL, SimGCL~\cite{yu2022graph} analyzes the reasons behind the performance improvement from contrastive learning and finds that the enhancement is largely independent of data augmentation. Instead, the performance improvement stems more from the contrastive learning loss enabling the model to learn more evenly distributed embeddings. XSimGCL~\cite{yu2023xsimgcl} further simplifies the SimGCL model.
GNN-based recommender models often suffer from over-smoothing, which causes user and item embeddings to become indistinguishably similar. To alleviate this issue, recent methods incorporate self-supervised and contrastive learning objectives to reshape the embedding distribution. Building upon SGL, SimGCL~\cite{yu2022graph} investigates the source of performance gains from contrastive learning and shows that the improvements are largely independent of data augmentation. Instead, the benefits primarily arise from the contrastive objective, which encourages a more uniformly distributed embedding space. XSimGCL~\cite{yu2023xsimgcl} further simplifies the SimGCL framework while maintaining its effectiveness.

\subsection{Sparse Models for Collaborative Filtering}
Sparse methods exploit user–item co-occurrence in the rating matrix to compute item-item similarities, enabling preference prediction without heavy parameter training. Sparse methods, such as ItemCF~\cite{linden2003amazon}, effectively use rating matrices to delve into item-item relationships, facilitating predictions without training. Similarly, slim~\cite{ning2011slim}, another traditional method, capitalizes on sparse structures, efficiently addressing collaborative filtering tasks~\cite{ning2011slim}. Beyond ItemCF~\cite{linden2003amazon} and slim~\cite{ning2011slim}, classic neighborhood methods include user-based kNN (UserCF~\cite{resnick1994grouplens}). More recently, EASE~\cite{steck2019embarrassingly} formulates item similarity as a sparse linear regression problem, achieving competitive accuracy with minimal training cost.

\subsection{Dual View Aligned Recommendation}

%Currently, some studies implicitly utilize both Sparse and Dense structures. Graph signal processing-based methods like GF-CF~\cite{shen2021powerful} employ sparse structures for modeling CF prediction tasks. GF-CF analyzes model design from the perspective of graph filtering, uniquely combining sparse item-item relationships with the dense properties of SVD~\cite{chen2021scalable}, resulting in remarkable model performance. UltraGCN~\cite{mao2021ultragcn} addresses the issue of over-smoothing by approximating an infinite-layer GNN and utilizes sparse item-item relationships to enhance the performance of graph neural networks. UltraGCN leverages sparse item-item relationships to assist Dense models in learning better embedding representations. PGSP~\cite{liu2023personalized} proposes a personalized graph signal processing framework that constructs mixed-frequency graph filters over enhanced similarity graphs, effectively blending local and global user signals to model sparse and dense information jointly.

Some recent studies implicitly leverage both sparse and dense structures in collaborative filtering. Graph signal processing-based methods, such as GF-CF~\cite{shen2021powerful}, model CF tasks using sparse item-item relationships while integrating dense properties from SVD~\cite{chen2021scalable}, yielding strong empirical performance. UltraGCN~\cite{mao2021ultragcn} mitigates over-smoothing by approximating an infinite-layer GNN and uses sparse item-item connections to guide dense embeddings, enhancing representation learning. PGSP~\cite{liu2023personalized} introduces a personalized graph signal processing framework that constructs mixed-frequency graph filters over enhanced similarity graphs, effectively combining local and global user signals to jointly capture sparse and dense information.

%Even though these methods demonstrate relatively good performance, their models are restricted to specific modules and are not plug-and-play. Moreover, their exploration of sparse and dense structures lacks an investigation into the underlying mechanisms, resulting in suboptimal performance and limited flexibility across recommendation scenarios. These limitations motivate the design of a generalizable, plug-and-play framework that fuses sparse and dense signals while enabling deep theoretical analysis of their complementary behaviors — a challenge this work addresses directly.

While these methods achieve competitive results, they are often limited to specific architectures or modules and lack a plug-and-play design. More importantly, they provide little theoretical insight into the complementary interactions between sparse and dense structures, which can lead to suboptimal performance and restricted flexibility across different recommendation scenarios. These limitations motivate the development of a generalizable, plug-and-play framework that not only fuses sparse and dense signals but also supports rigorous theoretical analysis of their synergistic behaviors — a challenge our work directly addresses.

%% file: Ours/chapter/preliminary.tex
\newcommand{\Ua}{\pazocal{U}}
\newcommand{\Va}{\pazocal{V}}
\newcommand{\Oa}{\pazocal{O}}
\newcommand{\Ya}{\pazocal{Y}}

\subsection{Problem Formulation}

In collaborative filtering, we denote the set of users as $\mathcal{U}$ ($|\mathcal{U}|=U$) and the set of items as $\mathcal{I}$ ($|\mathcal{I}|=I$). The foundational data is the interaction matrix $R \in \{0, 1\}^{U \times I}$, where an entry $r_{ui} = 1$ indicates an observed interaction between user $u$ and item $i$, and $r_{ui} = 0$ otherwise.

\paragraph{Dense View (Latent Embedding)} Dense models represent users and items as low-dimensional embeddings $E_U \in \mathbb{R}^{U \times d}$ and $E_I \in \mathbb{R}^{I \times d}$, capturing latent semantics in a continuous space. The primary goal of Dense models in this context is to predict the likelihood of a user \( u \) preferring an unobserved item \( i \). This likelihood, denoted as \( y^D_{ui} \), is typically computed using the dot product of the respective embeddings \( e_u \) and \( e_i \), encapsulated in Equation~\ref{eq:pre1}:
\begin{equation}
\label{eq:pre1}
y^D_{ui} = e_u^\top e_i.
\end{equation}

\paragraph{Sparse View (Explicit Structure)} Conversely, Sparse models operate directly on the observed interaction graph without mapping entities to latent spaces. They predict preference scores $y^S_{ui}$ by aggregating information from item neighborhoods. In matrix form, this is expressed as Equation~\ref{eq:pre2}:
\begin{equation}
\label{eq:pre2}
Y^S = RS,
\end{equation}
%where $S \in \mathbb{R}^{I \times I}$ is an item-item similarity (or weight) matrix. An entry $S_{ji}$ quantifies the contribution of item $j$'s historical interactions to the prediction of target item $i$. This paradigm relies on explicit co-occurrence patterns rather than latent feature abstractions. Table~\ref{tb:notation} presents the variables used in this paper.
where $S \in \mathbb{R}^{I \times I}$ denotes the item-item similarity (or weight) matrix. Each entry $S_{ji}$ measures the contribution of item $j$'s historical interactions to the prediction of the target item $i$. This approach relies on explicit co-occurrence patterns rather than latent feature representations. Table~\ref{tb:notation} summarizes the notations used throughout this paper.

\subsection{Theoretical Analysis}

In summary, this subsection establishes three key insights that inform the design of \textsc{SaD}: (i) dense models possess an attainable SNR ceiling on tail items; (ii) degree normalization offers only limited improvement and cannot break the $\sqrt{N}$ bottleneck; and (iii) introducing a complementary, weakly correlated sparse view enables SNR-improving fusion, thereby motivating our alignment choices.

We analyze why dense learning-based methods suffer on cold items from an SNR perspective, why degree normalization offers only limited relief, and how a structure-based view, together with a dual-view fusion, improves the attainable SNR. For a user–item positive/negative pair $(u,i^+;u,i^-)$, define the margin
$\Delta := y(u,i^+)-y(u,i^-)$ and,
\begin{equation}
\mathrm{SNR} := \frac{\mathbb{E}[\Delta]}{\sqrt{\operatorname{Var}(\Delta)}}.
\label{eq:snr-def}
\end{equation}

We use the signal-to-noise ratio (SNR) to measure how reliably a model can distinguish positive items from negatives — higher SNR indicates more stable ranking signals and less sensitivity to stochastic noise in training or sampling.

\subsubsection{SNR-based diagnosis}

This subsection diagnoses the intrinsic limitation (“SNR ceiling”) of dense models on tail items and motivates the need for a complementary, weakly correlated view to achieve effective fusion.

Consider MF with BCE loss, $z_{ui}=e_u^\top e_i$, $\sigma(z)=\tfrac{1}{1+e^{-z}}$.
Let the learned item embedding be $\hat e_i=e_i^\ast+\varepsilon_i$, with $\mathbb{E}[\varepsilon_i]=0$ and assume the covariance of the estimation noise follows the usual efficiency scaling:
\begin{equation}
\mathrm{Cov}(\varepsilon_i)\ \succeq\ \frac{c}{N_i}\,I_d \quad (c>0),
\label{eq:cov-scaling}
\end{equation}
where $N_i$ is the number of interactions of item $i$. This relation indicates that the estimation noise decreases proportionally to $1/N_i$—the more interactions an item has, the smaller its embedding variance. Since such noise directly affects the score difference between positive and negative items, it sets an upper bound on the achievable signal-to-noise ratio (SNR) of the model. 
Once the uncertainty of each item embedding is characterized, we can derive the margin for an item pair $(i,j)$ as:
\begin{equation}
\Delta^{\mathrm{MF}}_{ij}
= e_u^\top\!\big[(e_i^\ast-e_j^\ast)+(\varepsilon_i-\varepsilon_j)\big].
\end{equation}

\textbf{Theorem 1} (Cold Items Have a Lower Attainable SNR).

Under the conditions of Equation~\eqref{eq:cov-scaling} and $\|e_u\|\le B$, the signal-to-noise ratio (SNR) is bounded as follows:
\begin{equation}
\mathrm{SNR}_{\mathrm{MF}}(u,i,j)
\le
\frac{\|e_i^\ast-e_j^\ast\|}{\sqrt{c}}\;\sqrt{N_i}.
\label{eq:snr-mf-upper}
\end{equation}
Here, $c>0$ is a model-dependent constant controlling the noise scale in embedding estimation. Thus, tail items (small $N_i$) have a low SNR ceiling, making ranking unstable.

\textbf{Theorem 2} (Degree Normalization Cannot Break the $\sqrt{N_i}$ Bottleneck).

Degree normalization is a widely used heuristic to stabilize training by scaling scores according to user and item degrees. 
Specifically, it rescales the interaction score as
$\hat z_{ui}=\frac{e_u^\top e_i}{\sqrt{d_u d_i}}$.
This modifies the margin between items $i$ and $j$ to
\begin{equation}
\Delta^{\mathrm{norm}}_{ij}
=\frac{1}{\sqrt{d_u}}\Bigl(\frac{e_u^\top e_i}{\sqrt{d_i}}-\frac{e_u^\top e_j}{\sqrt{d_j}}\Bigr).
\end{equation}
Substituting $e_i=e_i^\ast+\varepsilon_i$ and using \eqref{eq:cov-scaling}, we obtain:
\begin{equation}
\mu_{\mathrm{norm}}
=\frac{1}{\sqrt{d_u}}\Bigl(\frac{e_u^\top e_i^\ast}{\sqrt{d_i}}-\frac{e_u^\top e_j^\ast}{\sqrt{d_j}}\Bigr),\quad
\sigma_{\mathrm{norm}}^2
\ \gtrsim\ 
\frac{\|e_u\|^2}{d_u}\Bigl(\frac{c}{d_i N_i}+\frac{c}{d_j N_j}\Bigr).
\end{equation}

Consequently, the item-dependent SNR factor still scales as
\begin{equation}
\frac{(1/\sqrt{d_i})}{\sqrt{1/(d_i N_i)}} = \sqrt{N_i},
\end{equation}
which is identical to the unnormalized case up to a constant multiplier. 
Therefore, while degree normalization may balance scale differences across users or items, 
it cannot fundamentally overcome the $\sqrt{N_i}$ statistical efficiency limit for cold items.

\textbf{Theorem 3} (Fusion Benefits Require Complementary and Weakly Correlated Views).

Let dense/sparse margins be $\Delta_1,\Delta_2$ with means $\mu_1,\mu_2$, stds $\sigma_1,\sigma_2$, and error correlation
$\rho=\mathrm{Corr}(\Delta_1-\mu_1,\Delta_2-\mu_2)\in[0,1)$.
For a convex blend $s=\alpha s_1+(1-\alpha)s_2$,
\begin{equation}
\mathrm{SNR}(\alpha,\rho)=
\frac{\alpha\mu_1+(1-\alpha)\mu_2}
{\sqrt{\alpha^2\sigma_1^2+(1-\alpha)^2\sigma_2^2+2\alpha(1-\alpha)\rho\,\sigma_1\sigma_2}}.
\label{eq:snr-fusion}
\end{equation}
Writing $r_k=\mu_k/\sigma_k$, we have $\frac{\partial}{\partial \rho}\mathrm{SNR}(\alpha,\rho)<0$ for any fixed $\alpha\in(0,1)$ with positive numerator, and
\begin{equation}
\exists\,\alpha:\ \mathrm{SNR}(\alpha,\rho)\ge\max\{r_1,r_2\}
\quad\Longleftrightarrow\quad
\rho\le\frac{r_{\min}}{r_{\max}},
\label{eq:rho-threshold}
\end{equation}
where $r_{\max}=\max\{r_1,r_2\}$, $r_{\min}=\min\{r_1,r_2\}$.
Moreover, for any $\alpha\in(0,1)$, it always holds that
\begin{equation}
\mathrm{SNR}(\alpha,\rho)\ge\min\{r_1,r_2\},
\end{equation}
indicating that convex fusion never underperforms the weaker view, and low cross-view correlation is crucial for surpassing the stronger one.

\subsubsection{Sparse view: neighborhood aggregation reduces variance (tail robustness)}

The above theoretical insights motivate the incorporation of a Sparse View, which serves as a low-correlation complement to the dense model and can effectively improve the overall signal-to-noise ratio (SNR). 
Next, we further analyze the additional advantages brought by the Sparse model. Structure-based methods (e.g., ItemCF/slim) aggregate local neighborhoods:
\begin{equation}
\hat f_{\mathrm{CF}}(u,i)=\sum_{j\in\mathcal{N}(i)} w_{ij}\,r_{uj}.
\end{equation}
Under local smoothness and $K_i:=|\mathcal{N}(i)|$ effective neighbors,
\begin{equation}
\operatorname{Var}(\hat f_{\mathrm{CF}}(u,i))\ \propto\ \frac{1}{K_i},
\label{eq:sparse-variance}
\end{equation}
so even when $N_i$ is small, the \emph{local} SNR can be high if $K_i$ is sufficient—explaining the robustness of structure-based models on tail items and their complementarity to dense methods.

\subsubsection{Further analysis: Dual-view fusion}

Building upon the naive dual-view strategy, we further analyze the potential for improvement and provide theoretical support for subsequent model design.

\textbf{Theorem 4} (Dual-View Fusion: Per-View SNR Gains lead to Higher Overall SNR). 
In a dual-view setting, improving the SNR of each individual view naturally enhances the overall fusion SNR.

\emph{Sparse $\Rightarrow$ Dense (pseudo-labels).}
When the sparse view provides additional pseudo-interactions for the dense model, it effectively enlarges the sample set of each item. Suppose $\eta_i$ denotes the number of these effective pseudo-interactions for item $i$. 
Then the covariance of the embedding noise decreases as:
\begin{equation}
\mathrm{Cov}(\varepsilon_i^{(D)\prime})\ \succeq\ \frac{c_D}{N_i+\eta_i}I_d,\qquad
\sigma_D'\ \approx\ \sigma_D\sqrt{\frac{N_i}{N_i+\eta_i}},
\label{eq:dense-var-reduce}
\end{equation}
and under label noise rate $\varepsilon<\tfrac12$,
\begin{equation}
\mu_D'\ \ge\ (1-2\varepsilon)\,\mu_D.
\label{eq:dense-mean-lb}
\end{equation}
Hence
\begin{equation}
r_D'=\frac{\mu_D'}{\sigma_D'}\ \gtrsim\ (1-2\varepsilon)\,r_D\,\sqrt{1+\frac{\eta_i}{N_i}}.
\end{equation}

\emph{Dense $\Rightarrow$ Sparse (item–item relations).}
Conversely, the dense model can enhance the sparse view by providing additional structural information through item–item relations. 
Let $\kappa_i$ denote the number of effective extra neighbors inferred from the dense embeddings for item $i$. 
These additional neighbors expand the aggregation set of the sparse model, thereby reducing the estimation variance:

\begin{equation}
\sigma_S'\ \approx\ \sigma_S\sqrt{\frac{K_i}{K_i+\kappa_i}},\qquad
\mu_S'\ \gtrsim\ \mu_S\ (\text{no harmful bias}),
\label{eq:sparse-var-reduce}
\end{equation}
thus
\begin{equation}
r_S'=\frac{\mu_S'}{\sigma_S'}\ \gtrsim\ r_S\,\sqrt{1+\frac{\kappa_i}{K_i}}.
\end{equation}

\paragraph{Monotonicity (no need to further reduce $\rho$).}
For fixed $\rho$, \eqref{eq:snr-fusion} is monotone increasing in each $\mu_k$ and decreasing in each $\sigma_k$.
If $\mu_k'\!\ge\!\mu_k$ and $\sigma_k'\!\le\!\sigma_k$ for $k=1,2$ with at least one strict, then
\begin{equation}
\max_{\alpha\in[0,1]}\mathrm{SNR}(\alpha,\rho)\ \text{with }(\mu',\sigma')
\ \ge\
\max_{\alpha\in[0,1]}\mathrm{SNR}(\alpha,\rho)\ \text{with }(\mu,\sigma).
\end{equation}
Thus, per-view SNR gains alone guarantee a higher fusion optimum, without requiring $\rho$ to decrease.

\paragraph{Trade-off when $\rho$ may increase.}
{Define the fusion envelope as:}
\begin{equation}
G(r_1,r_2,\rho)\ :=\ \sqrt{\frac{r_1^2+r_2^2-2\rho r_1 r_2}{1-\rho^2}}
\ =\ \sqrt{\,m^\top\Sigma^{-1}m\,}.
\label{eq:fusion-envelope}
\end{equation}
For small changes $(\Delta r_1,\Delta r_2,\Delta\rho)$ (allowing $\Delta\rho>0$),
\begin{equation}
(r_1-\rho r_2)\Delta r_1+(r_2-\rho r_1)\Delta r_2
\ \ge\ \frac{(r_1-\rho r_2)(r_2-\rho r_1)}{1-\rho^2}\,\Delta\rho
\ \Longrightarrow\ G\ \text{increases}.
\label{eq:tradeoff}
\end{equation}

Hence even if knowledge transfer slightly increases correlation, fusion SNR still improves provided per-view SNR gains are sufficiently large—precisely what \eqref{eq:dense-var-reduce}–\eqref{eq:sparse-var-reduce} deliver.

%% file: Ours/chapter/methods.tex
\newcommand{\Ha}{\pazocal{H}}
\newcommand{\Wa}{\pazocal{W}}
\newcommand{\La}{\pazocal{L}}
\newcommand{\Da}{\pazocal{D}}
\newcommand{\Qa}{\pazocal{Q}}

\subsection{Model Structure}

\begin{table}[h]
    \caption{Notations Table of SaD model}
    \label{tb:notation}
    \begin{tabularx}{0.9\textwidth}{p{0.15\textwidth}X}
        \Xhline{1pt}
         \textbf{Notation} & {\textbf{Definition}} \\
        \Xhline{0.5pt}
        % \addlinespace
        % \multicolumn{2}{l}{\small \textbf{Notations Related to Electronic Health Records (EHRs)}} \\
        % \Xhline{0.5pt}
        
        $\mathbf{E}_U $ & Embedding of User Group $\mathcal{U}$ \\ 
        $\mathbf{E}_I $ & Embedding of Item Group $\mathcal{I}$ \\ 
        $\mathbf{E}_u, \mathbf{E}_i$ & Embedding of a specific user or item\\ 
        $\mathbf{Y}^S$& Prediction results of the sparse model \\
        $\mathbf{Y}^D$ & Prediction matrix from the dense model \\
        $y_{ui}^S$& Prediction results of the sparse model for specific user and item\\
        $\mathbf{R}$ & Rating matrix \\ 
        $r_{ui}$ & Rating matrix element for $\mathbf{R}$ \\ 
        $\mathbf{S}$ & The similarity matrix of item-item relationship \\
        $s_{ij}$& The element of $\mathbf{S}$ \\
        $d_{ij}$& The variable used for normalization \\
        $\mathbf{R^*}$ &  Inferred rating matrix from the sparse model (used to augment dense view)\\
        $\mathbf{\hat{R}}$ & The rating matrix for dense after alignment \\
        $\mathbf{Q}$ & Inferred rating matrix from the dense model (used to augment sparse view) \\
        $\mathbf{R'}$ & The rating matrix for sparse after alignment \\
        $\widetilde{y}_{ui}$ & The final prediction results \\
        $\mathcal{U}$ & Set of users \\
        $\mathcal{I}$ & Set of items \\
        % \addlinespace
        % \multicolumn{2}{l}{\small \textbf{Notations Related to Hypergraphs}} \\  
        % \Xhline{0.5pt}

        \Xhline{1pt}
    \end{tabularx}
\end{table}

%From the above observation, we find that sparse-based methods and dense-based methods respectively exhibit good recommendation performance for unpopular and popular items. This observation suggests that these two paradigms are fundamentally \textbf{complementary}. A unified framework that synergizes these distinct advantages holds the promise of breaking the performance trade-off, enabling superior predictions across the entire item spectrum.

%Models leveraging Sparse structures and those utilizing Dense structures exhibit unique characteristics. By combining these two, we can enhance the model's performance substantially. In this section, we introduce the Sparse and Dense model framework, designed to incorporate both Sparse and Dense models in a flexible manner. Concurrently, we propose mechanisms for information transfer from Sparse to Dense and from Dense to Sparse. 

From the above observations, we note that \textbf{sparse-based methods} excel at recommending unpopular items, 
while \textbf{dense-based methods} perform well on popular items. 
This indicates that the two paradigms are fundamentally \textbf{complementary}. 
A unified framework that leverages the strengths of both approaches has the potential to overcome the performance trade-off, 
enabling superior predictions across the entire item spectrum.

Sparse- and dense-structure models exhibit distinct characteristics. 
By integrating these two perspectives, we can significantly enhance overall recommendation performance. 
In this section, we introduce the \textbf{Sparse and Dense (SaD) framework}, 
which flexibly incorporates both sparse and dense models. 
We also propose mechanisms for cross-view information transfer, 
allowing knowledge to flow from sparse to dense models and vice versa.

\subsubsection{Alignment Paradigm}

In our alignment paradigm, Dense models based on embeddings yield user embeddings \( \mathbf{E}_U \) and item embeddings \( \mathbf{E}_I \) as the final training output. For models utilizing SVD decomposition, low-rank factorization results in user features \( \mathbf{E}_U \) and item features \( \mathbf{E}_I \). By multiplying \( \mathbf{E}_U \) and \( \mathbf{E}_I \), we can obtain the output of the Dense model. In contrast, Sparse models typically directly produce the rating matrix \( \mathbf{Y}^S \in \mathbb{R}^{U \times I} \).

\subsubsection{Model Alignment}

The Sparse model fundamentally relies on the observed User-Item ($U-I$) interaction matrix and a learned Item-Item similarity matrix. The latter encapsulates explicit item correlations, providing high-confidence structural signals that can supplement the representation learning of the Dense model. Conversely, the Sparse model is inherently limited by the sparsity of the raw interaction matrix, often failing to capture unobserved potential interests. By integrating semantic insights from the Dense model, we can effectively enrich the input for the Sparse model, thereby enhancing its performance.

This integrated approach aims to leverage the distinct advantages of both views, creating a robust framework that effectively captures both the explicit structure in the data and the detailed latent features of users and items.

\begin{figure*}
    \centering
    \includegraphics[width=\linewidth]{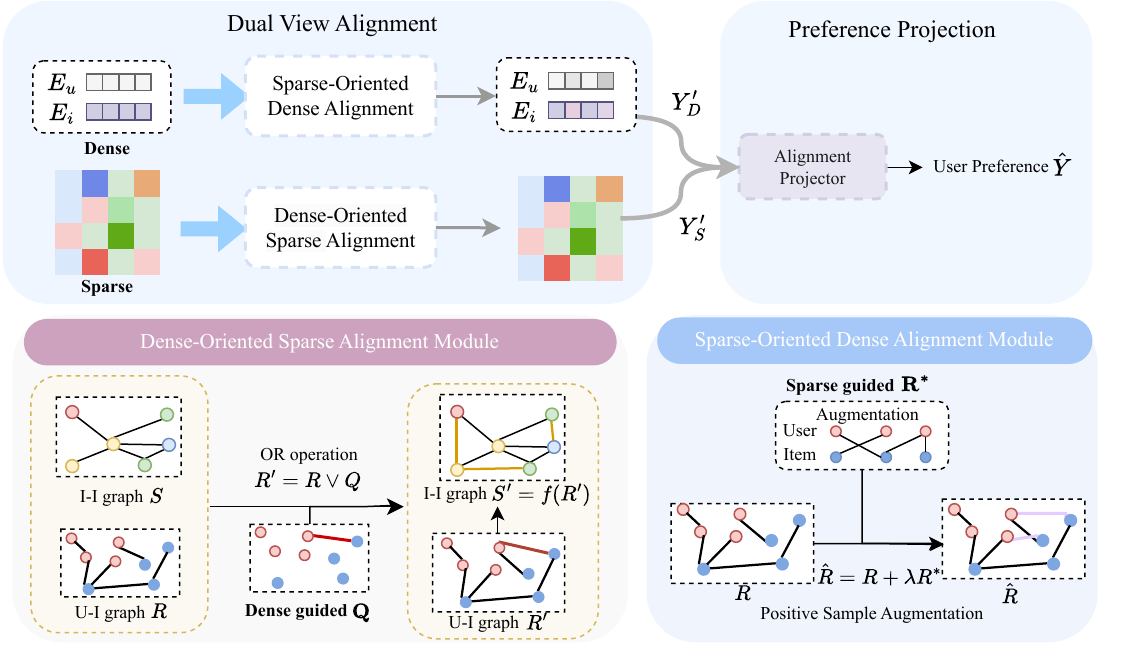}
    \caption{%SaD Model Structure Diagram. While performing within-view computations for both Dense and Sparse views, there is also an interaction alignment between the two views. Through the Sparse View and Dense View Alignment Module, each view obtained additional information from the other view. Finally, by projecting the two views via the Alignment Projector, the model incorporated the advantages of both views.
    Overview of the SaD model architecture. While each view—Dense and Sparse—performs its own computations, the two views interact through a cross-view alignment mechanism. The Sparse and Dense Alignment Module allows each view to incorporate complementary information from the other. Finally, the Alignment Projector integrates both views, enabling the model to leverage the strengths of each.}
    \label{fig:1}
\end{figure*}

\subsection{SaD Model}

Building upon the framework established in the previous section,  we implement two straightforward yet fundamentally robust models to represent the Sparse and Dense components, and facilitate interactions between them. We introduce a novel model, termed \textbf{SaD (Sparse and Dense)}, that effectively combines these components. 
%Our discussion will proceed with a detailed exploration of both the Sparse and Dense models, along with the mechanisms of information exchange between them. A visual representation of this model's architecture can be found in Figure~\ref{fig:1}, illustrating its intricate design.
In the following, we provide a detailed description of both the Sparse and Dense models, as well as the mechanisms that facilitate cross-view information exchange. 
A schematic overview of the model architecture is presented in Figure~\ref{fig:1}, highlighting its design and interactions.

\subsubsection{Sparse Module}

We employ slim (Sparse Linear Methods)~\cite{ning2011slim} as the backbone for the Sparse View due to its proven efficacy in capturing explicit item-item correlations. slim learns a sparse item-item similarity matrix \( \mathbf{S} \in \mathbb{R}^{I \times I} \), where the predicted interaction between user \( u \) and item \( i \) is estimated by linearly combining the user’s past interactions:

\begin{equation}
\label{eq:01}
y_{ui}^S = \sum_{j \in \mathcal{I}_u} r_{uj} \cdot s_{ji},
\end{equation}

\noindent where \( \mathcal{I}_u \) denotes the set of items interacted by user \( u \), and \( s_{ji} \) reflects the learned similarity between items \( j \) and \( i \).

The similarity matrix \( \mathbf{S} \) is learned by solving a regularized linear regression problem for each column \( s_i \) of \( \mathbf{S} \):

\begin{equation}
\label{eq:slim-loss}
\min_{s_i} \frac{1}{2} \|\mathbf{r}_i - \mathbf{R} s_i\|^2_2 + \lambda_1 \|s_i\|_1 + \lambda_2 \|s_i\|_2^2, \quad \text{s.t. } s_{ii} = 0,
\end{equation}

\noindent where \( \mathbf{r}_i \) is the interaction vector of item \( i \) and \( \mathbf{R} \in \mathbb{R}^{U \times I} \) is the user-item interaction matrix. The \( \ell_1 \) norm promotes sparsity, and the zero-diagonal constraint avoids trivial self-recommendations.

Once \( \mathbf{S} \) is learned, the prediction matrix \( \mathbf{Y}^S \in \mathbb{R}^{U \times I} \) is obtained by:

\begin{equation}
\label{eq:02}
\mathbf{Y}^S = \mathbf{R} \cdot \mathbf{S}.
\end{equation}

In the SaD framework, this module provides high-confidence local structural signals, which serve as a \textbf{source of structural supervision} for the Dense module via data augmentation during the alignment phase.

\subsubsection{Dense Module}

For the Dense View, we utilize a generalized Matrix Factorization (MF) framework tailored for implicit feedback. Users and items are mapped to latent embeddings $\mathbf{e}_u, \mathbf{e}_i \in \mathbb{R}^d$.

To stabilize training and mitigate popularity bias, we incorporate a graph-inspired normalization factor \( d_{ui} = \frac{1}{\sqrt{D_u D_i}} + \alpha \), where \( D_u \) and \( D_i \) denote the degree of user and item respectively in the interaction graph. The complete training loss is given by:

\begin{equation}
\begin{aligned}
   L_O = &-\sum \limits_{(u,i) \in \mathcal{N^{+}}} d_{ui} \log(\sigma(\mathbf{e}_u^\top \mathbf{e}_i)) \\
         &-\sum \limits_{(u,i) \in \mathcal{N^{-}}} d_{ui} \log(1 - \sigma(\mathbf{e}_u^\top \mathbf{e}_i)).
\end{aligned}
\label{eq:04}
\end{equation}

This formulation produces generalized latent representations, capturing \textbf{global semantic patterns} that complement the sparse view. Meanwhile, it is worth noting that SaD is model-agnostic; MF can be seamlessly replaced by advanced GNNs (e.g., LightGCN, SGL) without altering the alignment mechanism.

\subsection{Bidirectional Alignment Paradigm}

\subsubsection{Sparse-Oriented Dense View Alignment}

We modify the number of positive samples during training by influencing the Dense model's rating matrix $\mathbf{R}$. Given the sparsity of the user-item interaction matrix, we enhance its representational power by integrating the top-K nearest neighbors for items as determined by the Sparse model, resulting in an updated matrix \( \mathbf{\hat{R}} \).

The Sparse model's item-item relationship matrix is a key factor in transforming the user-item matrix. This process involves multiplying the Sparse model's item-item matrix with the original user-item matrix, followed by refining this enhanced matrix by selecting the top K neighbors based on the highest values for each user and item. This process effectively broadens the sparse user-item matrix.

\begin{equation}
\label{eq:05}
\begin{aligned}
    \text{Index}_u &= \text{TopK}_{i \in \mathcal{I}} \,y_{ui}^S, \\
    \text{Index}_i &= \text{TopK}_{u \in \mathcal{U}} \, y_{ui}^S.
\end{aligned}
\end{equation}

We define the pseudo-positive matrix $\mathbf{R}^*$ as Equation~\ref{eq:06}:
\begin{equation}
\label{eq:06}
r^*_{ui} = \mathbb{I}\left[u \in \text{Index}_i \, \lor \, i \in \text{Index}_u\right].
\end{equation}

Here, $\mathbb{I}[\cdot]$ denotes the indicator function, which returns 1 if the condition holds, and 0 otherwise.

We then incorporate these newly inferred samples (pseudo-positive) into the original rating matrix by treating them as a supplementary matrix $\mathbf{R^*}$, weighted by a confidence factor $\lambda \in [0, 1]$. The final enhanced rating matrix is given by $\hat{\mathbf{R}} = \mathbf{R} + \lambda \mathbf{R^*}$. This augmentation enriches the training signals for the Dense model, enabling it to learn more robust user-item representations.

For the newly added pseudo-positive interactions, the MF formula still applies. Considering these samples are added due to their relationship with neighboring items, the normalization matrix used is \( d_{ui} = \frac{1}{\sqrt{D_{U}D_I}} + \alpha I \). Here, \( D_{U} \) and \( D_I \) represent the degrees of users and items in the user-item matrix after adding new sample pairs. Assuming the set of new sample pairs is \( \mathcal{N'^+} \) and the corresponding negative sampling pairs are \( \mathcal{N'^-} \). Both positive and negative samples are obtained through sampling from the expanded matrix $\mathbf{\hat{R}}$. 

\begin{equation}
\label{eq:08}
\begin{aligned}   
   L_I=-\sum \limits_{(u,i)\in \mathcal{N'^{+}}} d_{ui}\log(\sigma(\mathbf{e}_u^T\mathbf{e}_i)) \\
   -\sum \limits_{(u,i)\in \mathcal{N'^-}}d_{ui}\log(1-\sigma(\mathbf{e}_u^T\mathbf{e}_i)).
\end{aligned}
\end{equation}

The \( L_I \) function represents the new training function. It is computed similarly to the original loss function but with an expanded set of training samples.

By doing this, we increase the number of training samples without modifying the fundamental structure of the model.

\subsubsection{Dense-Oriented Sparse View Alignment}

In sparse-view models, item-item similarity matrices are typically constructed by retaining only the top-K neighbors with the highest similarity scores, leading to a sparse representation. However, such sparsity may miss informative interactions. To address this, we leverage the Dense model to enrich the Sparse model by distilling high-confidence interactions.

After training the Dense model, we obtain user and item embeddings \( \mathbf{e}_u \) and \( \mathbf{e}_i \), as well as the predicted score matrix \( \mathbf{Y}^D \in \mathbb{R}^{U \times I} \). These outputs serve as a "teacher" to guide the construction of a more informative User-Item interaction matrix for the Sparse model.

Specifically, for each user, we select the top-K items with the highest scores in \( \mathbf{Y}^D \), and similarly, the top-K users for each item. The resulting matrix \( \mathbf{Q} \) represents pseudo-positive interactions derived from the Dense view. We then combine \( \mathbf{Q} \) with the original interaction matrix \( \mathbf{R} \) via an element-wise logical OR to form an augmented matrix:

\begin{equation}
\label{eq:09}
    \mathbf{R}' = \mathbf{R} \lor \mathbf{Q},
\end{equation}

\noindent where \( \lor \) denotes the element-wise logical OR. This enhanced matrix \( \mathbf{R}' \) captures both observed and inferred high-confidence interactions.

The Sparse model is retrained on \( \mathbf{R}' \), producing a refined similarity matrix \( \mathbf{S}' \). The final predictions are then computed as:

\begin{equation}
\label{eq:010}
    \mathbf{Y}'^S = \mathbf{R}' \cdot \mathbf{S}'.
\end{equation}

Through this dense-to-sparse alignment, we infuse global semantic knowledge from embeddings into the Sparse model, enabling it to better capture meaningful relationships under high sparsity. This bidirectional design enhances the robustness and overall accuracy of the SaD framework.

\subsubsection{Prediction}

Following the bidirectional view alignment between the Dense and Sparse models, we arrive at the stage of making the final prediction. By leveraging the Alignment Projector, the model combines the outputs of both the Dense and Sparse models through a weighted voting mechanism. This allows the model to integrate their strengths, ensuring a more robust and accurate predictive outcome.

\begin{equation}
\label{eq:011}
    \widetilde{y}_{ui} = \mathbf{W} \left( \begin{bmatrix} y_{ui}^D \\ y_{ui}^S \end{bmatrix} \right),
\end{equation}
where \( \mathbf{W} \in \mathbb{R}^{1 \times 2} \) is a projector that linearly combines the predictions from both models, \( y_{ui}' \) and \( p_{ui}' \), into a single unified output. Here, the matrix \( \mathbf{W} = \begin{bmatrix} 1,\beta \end{bmatrix} \) effectively scales and weights each input model's output, thereby capturing the balance between the Dense and Sparse contributions to generate the final prediction \( \widetilde{y}_{ui} \). This formulation highlights the projection operation, showcasing the transformation of the combined Dense and Sparse outcomes into a higher-level unified prediction through a projector. Specifically, the structure of the model is as outlined in Algorithm~\ref{alg:Framwork}.

\begin{algorithm}[htb]
\caption{SaD Model Process.}
\label{alg:Framwork}
\begin{algorithmic}[1] % This 1 denotes that each line is numbered
\REQUIRE ~~\\ % Algorithm input parameters: Input
    The user-item interaction rating matrix \( \mathbf{R} \in \mathbb{R}^{U \times I} \).\\
    The initial parameters of the model.
\ENSURE ~~\\ % Algorithm output: Output
    The final prediction result matrix \( \mathbf{Y} \in \mathbb{R}^{U \times I} \).

    \STATE Train and acquire the Sparse model according to Equation~\ref{eq:02};
    \STATE Enhance the Dense model based on the results from the Sparse model, deriving embeddings \( \mathbf{E}_U \) and \( \mathbf{E}_I \) following Equation~\ref{eq:08};
    \STATE Improve the Rating matrix with embeddings \( \mathbf{E}_U \) and \( \mathbf{E}_I \), then retrain the Sparse model in line with Equation~\ref{eq:010};
    \STATE Merge the models' results to generate the final prediction outcomes, using Equation~\ref{eq:011};
\end{algorithmic}
\end{algorithm}

\subsection{Discussion}

\subsubsection{Time Complexity Analysis}

%In our SaD model, the training time is inherently linked to the complexities of the Sparse (Slim) and Dense (MF) models. Specifically, the framework necessitates two computations of the Sparse (Slim) model and one computation of our Dense (MF) model. The total training time is the cumulative sum of these individual training times.

In our \textsc{SaD} model, the training time is primarily determined by the complexities of the Sparse (slim) and Dense (MF) components. Specifically, the framework requires two forward passes of the Sparse (slim) model and one forward pass of the Dense (MF) model. Consequently, the total training time is the sum of these individual computations.

%For our framework, this equates to twice the training time of the Slim model and once for the MF model. The overall complexity for our framework can be approximated as \( 2 \times T_{Slim} + T_{MF} \). The inclusion of new interactions in the Dense model from the Sparse model leads to faster convergence, potentially offsetting the increased computational demand.

Formally, the overall complexity can be approximated as \(2 \times T_{\mathrm{slim}} + T_{\mathrm{MF}}\). Importantly, incorporating information from the Sparse model into the Dense model accelerates convergence, which may partially offset the additional computational cost.

%The corresponding actual run times are shown in Table~\ref{tab:complexity_breakdown}. The Dense model is not the original MF model, which accounts for the difference in time.

The actual run times for each component are summarized in Table~\ref{tab:complexity_breakdown}. Note that the Dense model used here is not the original MF model, which explains the observed differences in computation time.

% \begin{table}[ht!]
%     \centering
%     \caption{The time comparison between SaD and other models on different datasets.}
%     \setlength{\tabcolsep}{12pt}
%     \renewcommand{\arraystretch}{1.2}
%     \begin{tabular}{c|cc|cc}
%     \toprule
%     Model & \multicolumn{2}{c}{Yelp}& \multicolumn{2}{c}{Moivelens} \\ 
%     \cline{2-5}
%     & Time cost& Recall & Time cost &Recall \\
%     \hline
%     LightGCN & 11 h 3 min&0.0649 & 3h53min&0.2576 \\
    
%     UltraGCN &45min&0.0683   &28min&0.2787  \\

%     SGL &  36min&0.0669&42min&0.2577  \\ 
    
%     SimGCL&29min&0.0724&27min&0.2711\\ \hline
%     SaD &  35 min&\textbf{0.0731} & 16min&\textbf{0.2865} \\
%     \bottomrule
%     \end{tabular}
%     \label{tab:table1}
% \end{table}

\begin{table}[ht!]
  \centering
  \caption{Detailed time comparison on Yelp and Movielens. Recall is reported only for full models.}
  \setlength{\tabcolsep}{8pt}
  \renewcommand{\arraystretch}{1.2}
  \begin{tabular}{l l cc cc}
    \toprule
    \multirow{2}{*}{\textbf{Model}} & \multirow{2}{*}{\textbf{Module}} & \multicolumn{2}{c}{\textbf{Yelp}} & \multicolumn{2}{c}{\textbf{Movielens}} \\
    \cmidrule(lr){3-4}\cmidrule(lr){5-6}
     &  & Time (h:min) & Recall@20 & Time (h:min) & Recall@20 \\
    \midrule
    LightGCN & Total & 11:03 & 0.0649 & 3:53 & 0.2576 \\
    UltraGCN & Total & 0:45  & 0.0683 & 0:28 & 0.2787 \\
    SGL      & Total & 0:36  & 0.0669 & 0:42 & 0.2577 \\
    SimGCL   & Total & 0:29  & 0.0724 & 0:27 & 0.2711 \\
    \midrule
    \multirow{4}{*}{SaD}
      & Total        & 0:35 & \textbf{0.0731} & 0:16 & \textbf{0.2865} \\
      & Dense Train  & 0:21 & —                & 0:11 & — \\
      & Sparse Train & 0:09 & —                & 0:02 & — \\
      & Alignment    & 0:05 & —                & 0:03 & — \\
    \bottomrule
  \end{tabular}
  \label{tab:complexity_breakdown}
\end{table}

\subsubsection{Relations with other Models}

\textbf{Flexibility with Dense Models.} 
The SaD framework is designed to be modular, meaning the Dense and Sparse components can be easily swapped. In this work, we use a basic Matrix Factorization (MF) as the dense model to show that our improvements come from the alignment method itself, not from using a complex backbone. However, the framework is highly flexible: the dense component can be replaced by advanced GNNs (e.g., LightGCN, SGL) or other models without changing the overall alignment structure.

\textbf{Complementarity with Self-Supervised Learning (SSL).} 
SaD works alongside Self-Supervised Learning rather than competing with it. Existing SSL methods (e.g., SGL, SimGCL) typically work within a single view (e.g., perturbing the graph) to improve feature quality. In contrast, SaD performs \textit{cross-view} alignment, using the Sparse view to fix missing information in the Dense view (and vice versa). These two approaches solve different problems. Therefore, combining SSL-based models with SaD leads to even better results, as shown in Section~\ref{sec:generalizability}.

%% file: Ours/chapter/experiments.tex
In this section, we detail our experimental setup and report the results. Our experiments are designed to systematically answer the following research questions:

\begin{itemize}
    \item \textbf{RQ1:} \emph{Overall Effectiveness.} Does the proposed model achieve superior recommendation performance compared to state-of-the-art baselines across diverse datasets?
    \item \textbf{RQ2:} \emph{Module Contribution.} What is the individual contribution of each key component in our framework, as revealed by ablation studies?
    \item \textbf{RQ3:} \emph{Long-tail Robustness.} Can our method effectively alleviate the performance degradation observed in Dense models for unpopular (long-tail) items?
    \item \textbf{RQ4:} \emph{Hyperparameter Sensitivity.} How sensitive is the model's performance to changes in key hyperparameters?
    \item \textbf{RQ5:} \emph{Plug-and-Play Capability.} Can our framework be seamlessly integrated into different backbone recommendation models, demonstrating strong plug-and-play usage?
\end{itemize}

This comprehensive experimental evaluation aims to validate the overall effectiveness, dissect the inner mechanisms, and demonstrate the robustness and flexibility of our approach under various scenarios.

\begin{table}[ht!]
  \centering
  \caption{Overview of the four datasets.}
  \label{tab:table3}
  \setlength{\tabcolsep}{12pt} % 调整列间距
  \renewcommand{\arraystretch}{1.2} % 调整行间距
  \begin{tabular}{l r r r r}
     \toprule
    Dataset & \multicolumn{1}{c}{User} & \multicolumn{1}{c}{Item} & \multicolumn{1}{c}{Interaction} & \multicolumn{1}{c}{Sparsity} \\
    \midrule
    Gowalla & 29,858 & 40,981 & 1,027,370 & 99.91\% \\
    Yelp2018 & 31,668 & 38,048 & 1,561,406 & 99.87\% \\
    Amazon-book & 52,643 & 91,599 & 2,984,108 & 99.94\% \\
    Movielens & 6,022 & 3,043 & 796,244 & 95.65\% \\
    \bottomrule
  \end{tabular}
\end{table}

\subsection{Experimental Settings}
%Four public datasets are used in the experiments: Yelp\footnote{\url{https://www.yelp.com/dataset}}, Gowalla\footnote{\url{http://snap.stanford.edu/data/loc-gowalla.html}}, Amazon-Book\footnote{\url{https://jmcauley.ucsd.edu/data/amazon/}}, and Movielens\footnote{\url{https://grouplens.org/datasets/movielens/}}. Following the division methods commonly employed in the field of Collaborative Filtering, we conduct experiments under the same data partition as \cite{he2020lightgcn}. Our data partitioning method on Amazon-book, Yelp, and Gowalla is also the same as that on the BarsMatch Leaderboard. These diverse datasets enable us to comprehensively assess the performance and applicability of our model. The user-item interaction information for these four datasets is illustrated in Table~\ref{tab:table3}.

We conduct experiments on four publicly available datasets: Yelp\footnote{\url{https://www.yelp.com/dataset}}, Gowalla\footnote{\url{http://snap.stanford.edu/data/loc-gowalla.html}}, Amazon-Book\footnote{\url{https://jmcauley.ucsd.edu/data/amazon/}}, and Movielens\footnote{\url{https://grouplens.org/datasets/movielens/}}. Following common practices in collaborative filtering research, we adopt the same data partitioning protocol as in \cite{he2020lightgcn}. For Amazon-Book, Yelp, and Gowalla, our partitioning is consistent with that used on the BarsMatch leaderboard. This selection of diverse datasets allows for a comprehensive evaluation of our model's performance and generalizability. Summary statistics of user-item interactions for the four datasets are provided in Table~\ref{tab:table3}.

\subsubsection{Baseline Models}

We compare our method with two categories of baselines. The first category includes strong general recommendation models, while the second category consists of methods that \emph{implicitly} use dual view designs. 

\textbf{General Recommendation Baselines:} This category includes LR-GCCF~\cite{chen2020revisiting}, ENMF~\cite{chen2020efficient}, NIA-GCN, LightGCN~\cite{he2020lightgcn}, SGL~\cite{wu2021self}, DGCF~\cite{liu2021graph}, and SimpleX~\cite{mao2021simplex}. These models cover mainstream GCN-based recommenders, matrix factorization improvements, and recent self-supervised or attention-based approaches. We briefly introduce several representative baselines below:

\begin{itemize}
    \item \textbf{LightGCN}~\cite{he2020lightgcn}: LightGCN is a highly efficient and influential graph neural network for recommendation. It removes feature transformation and nonlinear activation from classical GCNs, focusing solely on neighborhood aggregation to learn node embeddings. This lightweight architecture not only improves scalability but also achieves state-of-the-art performance on many benchmark datasets.

    \item \textbf{SGL}~\cite{wu2021self}: Self-supervised Graph Learning (SGL) is the first work to introduce self-supervised learning signals into GCN-based recommendation. SGL constructs contrastive tasks by perturbing the user-item interaction graph, encouraging the model to learn more robust representations, thereby boosting performance, especially in sparse scenarios.

    \item \textbf{SimpleX}~\cite{mao2021simplex}: SimpleX improves the classical Matrix Factorization (MF) framework by modifying the loss function and optimizing the negative sampling process. By introducing a self-supervised negative sampling strategy and a better loss design, SimpleX achieves better item ranking, particularly for cold items.

    \item \textbf{SimGCL}~\cite{yu2022graph}: SimGCL is a recent model that integrates contrastive learning with graph-based collaborative filtering. Unlike SGL, SimGCL introduces random noise into node embeddings before contrastive learning, effectively regularizing the model and improving both accuracy and robustness.
\end{itemize}

%\textbf{Implicit Dual View-based Methods:} Our main comparison is with methods that implicitly leverage dual-view modeling, including UltraGCN~\cite{mao2021ultragcn}, GF-CF~\cite{shen2021powerful}, and PGSP~\cite{liu2023personalized}. These three state-of-the-art models represent the leading performance on three widely recognized authoritative data partitions. All these models implicitly combine global structure and local similarity information, aligning closely with our research focus. The following provides a brief overview of these methods:
\textbf{Implicit Dual View-based Methods:} Our primary comparison focuses on methods that implicitly leverage dual-view modeling, including UltraGCN~\cite{mao2021ultragcn}, GF-CF~\cite{shen2021powerful}, and PGSP~\cite{liu2023personalized}. These state-of-the-art models achieve leading performance on widely recognized benchmark data partitions. Each method implicitly integrates global structural information with local similarity signals, closely aligning with the dual-view perspective central to our work. A brief overview of these methods is provided below:

\begin{itemize}
    \item \textbf{UltraGCN}~\cite{mao2021ultragcn}: UltraGCN is a scalable and efficient GCN-based recommendation model that mathematically approximates the effect of infinite-layer propagation in standard GCNs. It leverages a weighted combination of user and item features, allowing the model to capture long-range dependencies without deep stacking, and offers high efficiency for large-scale recommendation.

    \item \textbf{GF-CF}~\cite{shen2021powerful}: Graph Filtering for Collaborative Filtering (GF-CF) models collaborative filtering relationships via spectral graph filtering. By treating the user-item interaction matrix as a graph signal, GF-CF applies low-pass filters to extract smooth collaborative information, improving the generalization ability of the learned representations.

    \item \textbf{PGSP}~\cite{liu2023personalized}: Power Graph Signal Processing (PGSP) extends GF-CF by additionally modeling high-frequency components in the graph signal. PGSP utilizes both low-frequency (smooth) and high-frequency (personalized, local) information in the user-item graph, leading to improved recommendations, especially for long-tail and cold-start items.
\end{itemize}

\begin{table*}[]
\centering
\normalsize
%\caption{Comparison of performance on four datasets. To guarantee fairness in our comparison, all model results are as reported by prior research. Subsequent experiments will demonstrate the plug-and-play capability of our model, showing improved performance with common Dense models.
\caption{Performance comparison on four datasets. To ensure fairness, all baseline results are taken from prior studies. Subsequent experiments highlight the plug-and-play capability of our model, demonstrating consistent improvements when integrated with standard Dense models.}
\setlength{\tabcolsep}{3.5pt}
\label{tab:table2}
\renewcommand{\arraystretch}{1.2}
\begin{tabular}{c c c c c c c c c}
    \toprule
    Methods & \multicolumn{2}{c}{Yelp2018} & \multicolumn{2}{c}{Gowalla} & \multicolumn{2}{c}{Amazonbooks} & \multicolumn{2}{c}{Movielens}\\
    \cline{2-3} \cline{4-5} \cline{6-7} \cline{8-9}
     & Recall@20 & NDCG@20 & Recall@20 & NDCG@20 & Recall@20 & NDCG@20 & Recall@20 & NDCG@20\\
    \midrule
    LR-GCCF & 0.1519 & 0.1285 & 0.0561 & 0.0343 & 0.0335 & 0.0265& 0.2231 & 0.2124 \\
    ENMF & 0.1523 & 0.1315 & 0.0624 & 0.0515 & 0.0359 & 0.0281 & 0.2315 & 0.2069\\
    NIA-GCN & 0.1726 & 0.1358 & 0.0599 & 0.0491 & 0.0369 & 0.0287 & 0.2359 & 0.2242\\
    NGCF & 0.1570 & 0.1327 & 0.0579 & 0.0477 & 0.0344 & 0.0263 & 0.2513& 0.2511\\
    LightGCN & 0.1830 & 0.1554 & 0.0649 & 0.0530 & 0.0411 & 0.0315 & 0.2576 & 0.2427\\
    DGCF & 0.1842 & 0.1561 & 0.0654 & 0.0534 & 0.0422 & 0.0324 & 0.2640& 0.2504\\
    
    SGL-ED & 0.1769 & 0.1502 & 0.0675 & 0.0555 & 0.0478 & 0.0379& 0.2577 & 0.2507 \\
    
    % DGCF & 0.1891 & 0.1602 & 0.0703 & 0.0575 & 0.0476 & 0.0369 & - & -\\
    SimpleX & 0.1872 & 0.1557 & 0.0701 & 0.0575 & 0.0583 & 0.0468 & \underline{0.2802} & \underline{0.2670}\\
     SimGCL&0.1830&0.1542&\underline{0.0724}&\underline{0.0596}&0.0480&0.0377&0.2728&0.2594\\

     slim & 0.1699 & 0.1382 & 0.0646 & 0.0541 & \underline{0.0755} & \underline{0.0602} & 0.2577 & 0.2507\\
     
    \midrule
    GF-CF & 0.1849 & 0.1518 & 0.0697 & 0.0571 & 0.0710 & 0.0584 & 0.2667 & 0.2549\\
    UltraGCN & 0.1862 & 0.1580 & 0.0683 & 0.0561 & 0.0681 & 0.0556 &0.2787& 0.2642\\
    PGSP & \underline{0.1916} &\underline{0.1605} & 0.0712 &0.0587 & 0.0710& 0.0583 & 0.2759 & 0.2627 \\
    \midrule
    \rowcolor{gray!15} SaD       & \textbf{0.1969}        & \textbf{0.1652}           & \textbf{0.0731}         & \textbf{0.0602} & \textbf{0.0796}& \textbf{0.0635}  & \textbf{0.2865} & \textbf{0.2743}     \\
    \midrule

    \textit{Improves (\%)} & 
2.8\% & 2.9\% & 1.0\% & 1.0\% & 5.4\% & 5.5\% & 2.3\% & 2.7\% \\
\textit{p-value} & 
7.3e-7 & 3.0e-7 & 5.5e-3 & 4.3e-3 & 2.4e-9 & 6.5e-9 & 1.4e-5 & 7.1e-6 \\
\bottomrule
  \end{tabular}
\end{table*}

\subsubsection{Implement Details}

%In our experiments, we employ the same train-validation-test split strategy for the four datasets as utilized in previous studies~\cite{shen2021powerful,mao2021ultragcn,he2020lightgcn,mao2021simplex,wu2021self}. Our model is implemented using Pytorch. We evaluate its performance using the widely accepted collaborative filtering metrics Recall@20 and NDCG@20. To align with existing research, we utilize the all-items ranking method for evaluation. Our experiments are carried out on a V100 GPU. We use the same embedding size of 64 as in previous work. During the Dense to Sparse information transfer, the search range for the $K$ value in the top-K operation is set to [0, 5, 10, 15, 20, 25, 30, 50]. For the deep learning component, we set the learning rate at $1e^{-3}$, with a batch size of 1024 and an L2 regularization norm of $1e^{-4}$. The hyperparameter $\beta$, representing the combination weight between Dense and Sparse models, is varied across [1, 3, 5, 10, 15, 20, 50, 100, 200, $1e^3$, $1e^4$] to find the optimal setting.

In our experiments, we follow the same train-validation-test split strategy for all four datasets as in prior studies~\cite{shen2021powerful,mao2021ultragcn,he2020lightgcn,mao2021simplex,wu2021self}. Our model is implemented in PyTorch, and performance is evaluated using standard collaborative filtering metrics: Recall@20 and NDCG@20. Consistent with previous work, we adopt the all-items ranking protocol for evaluation. All experiments are conducted on a V100 GPU.

We use an embedding size of 64, matching prior studies. During Dense-to-Sparse information transfer, the search range for the top-$K$ operation is set to $K \in \{0,5,10,15,20,25,30,50\}$. For the deep learning components, we use a learning rate of $1\mathrm{e}{-3}$, a batch size of 1024, and an L2 regularization of $1\mathrm{e}{-4}$. The hyperparameter $\beta$, which controls the combination weight between Dense and Sparse models, is searched over $\{1, 3, 5, 10, 15, 20, 50, 100, 200, 10^3, 10^4\}$ to determine the optimal setting.

\subsection{Performance}

%The performance of SaD is presented in Table~\ref{tab:table2}. In this section, we will compare our model with previous works. As can be seen from the table, our model performance is currently the most advanced, surpassing all existing models in the collaborative filtering task of the recommendation system. The SaD model effectively enhances the model's performance by integrating Sparse structure and Dense information. For all the baseline models, we have adopted the results reported in prior research. 
The performance of SaD is summarized in Table~\ref{tab:table2}, where we compare our model against prior state-of-the-art methods. As shown, SaD achieves the highest performance across all evaluated datasets, demonstrating the effectiveness of integrating Sparse structures with Dense information. For all baseline models, we report the results as provided in the original publications to ensure a fair comparison.

%Our model achieved leading performance on four datasets, with particularly notable results on Gowalla, Yelp, and Amazon-book. Specifically, in terms of Recall, the model outperformed the best baseline model by 2.8\%, 1.0\%, and 5.4\% on the Gowalla, Yelp, and Amazon-book datasets, respectively.
Our model consistently outperforms existing methods on all four datasets, with particularly notable gains on Gowalla, Yelp, and Amazon-Book. In terms of Recall@20, SaD surpasses the best-performing baseline by 2.8\%, 1.0\%, and 5.4\% on Gowalla, Yelp, and Amazon-Book, respectively.

%It is worth noting that for the sake of clarity, we initially chose MF as the representative Dense model. Subsequent experiments have shown that replacing MF with more advanced Dense models, such as SimGCL, enhances the performance of our SaD framework significantly.
We emphasize that MF was initially chosen as the representative Dense model for clarity and simplicity. Subsequent experiments indicate that replacing MF with more advanced Dense models, such as SimGCL, further improves the performance of the SaD framework, highlighting its plug-and-play flexibility.

\subsection{Ablation Study}
%In this section, we conduct ablation experiments on the model, evaluating the impact of removing the Sparse module and Dense module on performance. Since our model consists of both the Dense and Sparse modules, we can assess the contribution of each module to the overall performance. Concurrently, we can also evaluate the contribution of the proposed information interaction mechanism to the model's performance.
In this section, we perform ablation experiments to evaluate the contribution of each component in the SaD model. Specifically, we examine the impact of removing the Sparse module or the Dense module on overall performance. Since SaD integrates both modules, these experiments allow us to quantify the individual contributions of each view. Additionally, we assess the effect of the proposed cross-view information interaction mechanism, highlighting its role in enhancing the model's predictive capabilities.

\begin{table*}[]
  \begin{center}
    \caption{
Performance of Different SaD Variants.}
    \setlength{\tabcolsep}{2pt}
    \label{tab:table4}
    \renewcommand{\arraystretch}{1.2}
    \begin{tabular}{ccccccccc}
    \toprule
      Methods & \multicolumn{2}{c}{Gowalla} & \multicolumn{2}{c}{Yelp2018} & \multicolumn{2}{c}{Amazonbooks} & \multicolumn{2}{c}{Movielens}\\
    \cline{2-3} \cline{4-5} \cline{6-7} \cline{8-9}
     & Recall@20 & NDCG@20 & Recall@20 & NDCG@20 & Recall@20 & NDCG@20& Recall@20 & NDCG@20\\
      \midrule
      Dense                         & 0.0674& 0.0557        & 0.1852         & 0.1562  &0.0499&0.0392 &0.2777&  0.2625    \\ 
Sparse                         & 0.0655         & 0.0545        & 0.1635        & 0.1322  &0.0755&0.0602 &0.2577&0.2507     \\ 
 Ensemble &0.0682 &0.0565 &0.1863        &0.1580 &0.0521&0.0414&0.2828 &0.2708\\
 SaD w/o $\rho$ analysis &0.0679 & 0.0559 & 0.1858  & 0.1568 & 0.0527 & 0.0417& 0.2756 & 0.2613\\
\midrule

Dense-R&0.0689&0.0567& 0.1870& 0.1576&0.0627& 0.0510&0.2798&0.2658\\
    Sparse-R&0.0685&0.0570& 0.1783  & 0.1466&0.0788&0.0628&0.2597&  0.2531\\
SaD w/o align&  0.0715&0.0591&0.1922&   0.1611&0.0765&0.0608&0.2847 &0.2737\\ 
    SaD w/o S2D  & 0.0723& 0.0599 & 0.1968& 0.1647 & 0.0779 &0.0620 & 0.2852 &0.2740 \\ 
      SaD w/o D2S  & 0.0719&0.0592&0.1924& 0.1611&0.0775&0.0617&0.2857&0.2741   \\ \midrule
\rowcolor{gray!15} SaD    & \textbf{0.0731}&\textbf{0.0602}&\textbf{0.1969}& \textbf{0.1652}&\textbf{0.0796}&\textbf{0.0635}&\textbf{0.2865}&\textbf{0.2743}       \\ \bottomrule
    
    \end{tabular}
    
  \end{center}
\end{table*}

In Table~\ref{tab:table4}, Dense and Sparse represent the standalone Dense and Sparse models without information transfer refinement. Dense-R and Sparse-R denote the Dense and Sparse models with information transfer refinement. Ensemble indicates the simple summation of Dense and Sparse. SaD w/o $\rho$ is a variant that directly combines two dense models (LightGCN and MF). Without the $\rho$-guided mechanism, it fuses highly correlated dense representations rather than exploiting complementary sparse information. SaD w/o align denotes the scenario without Dense-to-Sparse (D2S) and Sparse-to-Dense (S2D) Alignment. SaD w/o S2D means excluding the Sparse-to-Dense module, while SaD w/o D2S indicates the removal of the Dense-to-Sparse module.

%Table~\ref{tab:table4} shows that removing either view degrades performance, while combining both views with mutual enhancement consistently improves results. At the same time, augmenting the sample number of the Dense model through the Sparse model and enhancing the operations of the Sparse module through the Dense model can further improve the performance of the combined SaD model. We can observe that, by introducing Sparse models on top of Dense models and designing information distillation between the two views, we can significantly enhance the performance of Dense models. Meanwhile, the view alignment brings about an improvement of 2\%-4\% against SaD w/o align, further confirming that SaD’s advantage arises from effective cross-view interaction.

Table~\ref{tab:table4} demonstrates that removing either the Sparse or Dense view leads to a noticeable drop in performance, whereas integrating both views with mutual enhancement consistently yields superior results. Furthermore, augmenting the sample size for the Dense model using information from the Sparse model, and simultaneously enriching the Sparse module with Dense embeddings, further boosts the performance of the combined SaD framework. These observations indicate that introducing Sparse models on top of Dense models, along with the proposed cross-view information distillation, substantially enhances Dense model performance. In addition, the view alignment contributes an improvement of 2\%--4\% over SaD without alignment, confirming that SaD’s advantage stems from effective cross-view interaction.

\subsection{Benefits of SaD}

%Beyond performance, the prediction accuracy for unpopular items is crucial due to the presence of popularity bias.

%\textbf{Long-tail Recommendation} The collaborative filtering methods based on Graph Neural Networks (GNNs) often suffer from the issue of popularity bias in recommender systems. For items, the higher their popularity, the more likely they are to be recommended by the system~\cite{clauset2009power,tang2020investigating}. Conversely, long-tail items with fewer interactions have a lower probability of being recommended. However, our SaD method combines structured sparse techniques, effectively enhancing the recommendation performance for long-tail items. As depicted in the graph, our approach significantly improves the performance for long-tail items compared to the dense model.

Beyond overall performance, improving prediction accuracy for unpopular items is particularly important due to the inherent popularity bias in recommendation systems.

\textbf{Long-tail Recommendation.} Collaborative filtering methods based on Graph Neural Networks (GNNs) often exhibit a strong popularity bias: items with higher popularity are more likely to be recommended~\cite{clauset2009power,tang2020investigating}, while long-tail items with fewer interactions are underrepresented. In contrast, our SaD framework leverages structured sparse techniques to effectively enhance recommendations for long-tail items. As illustrated in the figure, SaD substantially improves the performance on long-tail items compared to conventional dense models.

%Here, we will refer to the division criteria and split the data into three groups: Unpopular, Normal, and Popular. The Unpopular group consists of the 80\% of items with the fewest interactions, the Normal group includes items with 80\% to 95\% of interactions, and the Popular group contains the top 5\% of items with the most interactions. We will conduct tests on this data split to evaluate our model's performance on unpopular items.

Here, we divide the items into three groups based on the number of user interactions: \textbf{Unpopular}, \textbf{Normal}, and \textbf{Popular}. The \textbf{Unpopular} group consists of the 80\% of items with the fewest interactions, the \textbf{Normal} group includes items ranked between the 80th and 95th percentiles, and the \textbf{Popular} group comprises the top 5\% of items with the most interactions. We evaluate our model’s performance specifically on these groups to assess its effectiveness on long-tail (unpopular) items.

\begin{figure}[h!]
  \centering
  \subfloat[Performance on Yelp]{%
    \includegraphics[width=0.48\linewidth]{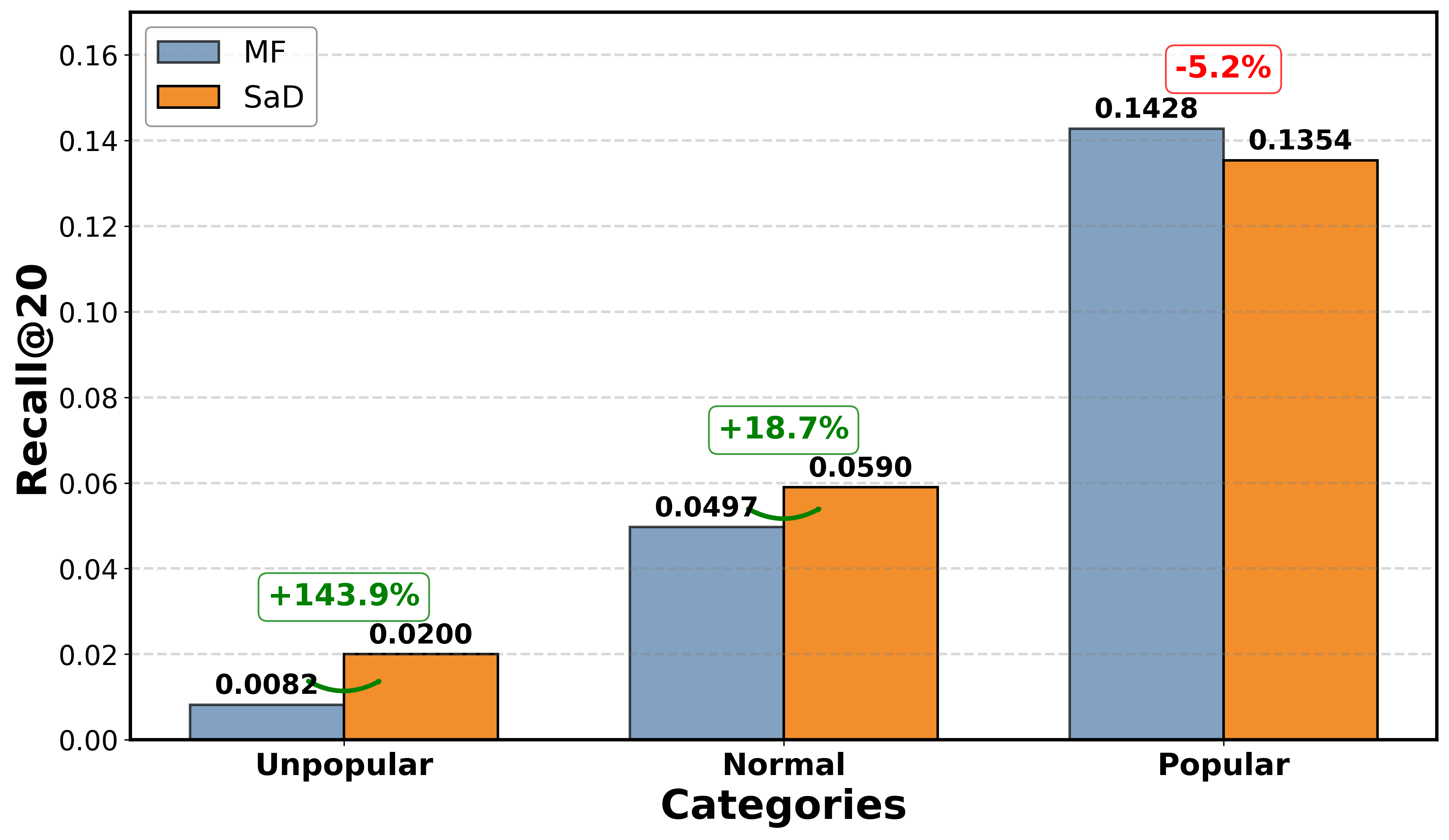}%
    \label{fig:image1}%
  }\hfill
  \subfloat[Performance on Movielens]{%
    \includegraphics[width=0.48\linewidth]{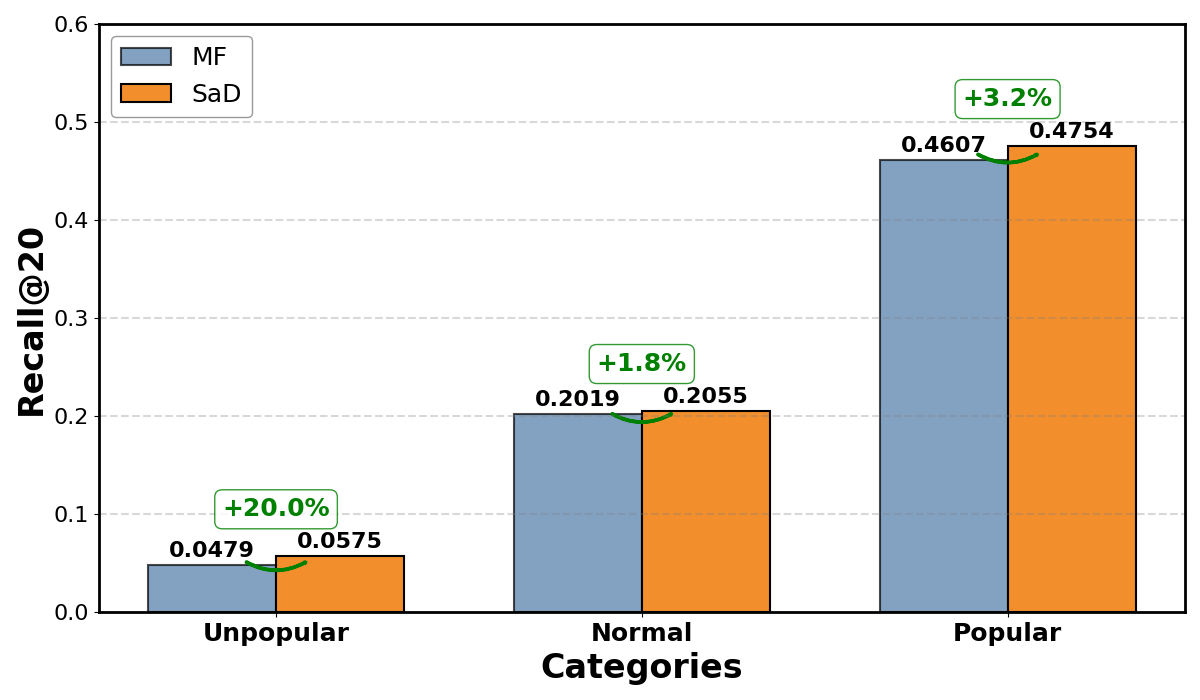}%
    \label{fig:image2}%
  }
  %\caption{Performance of different groups. Under our SaD framework, MF shows significant performance improvement on unpopular items.}
  \caption{Performance across item popularity groups. Under the SaD framework, the MF model demonstrates substantial improvements on unpopular (long-tail) items.}
  \label{fig:2}
\end{figure}

%Figure~\ref{fig:2} clearly shows that across various groups, our model consistently outperforms the Dense model. Notably, our model demonstrates significant performance improvements for less popular items. The performance improvement of the SaD model, compared to conventional deep learning-based models, is largely due to its enhanced performance on unpopular items. On the Movielens and Gowalla datasets, the SaD model significantly improves the performance on unpopular items. Specifically, on the Movielens dataset, the model achieves approximately a 25\% performance improvement for unpopular items.

Figure~\ref{fig:2} clearly illustrates that our model consistently outperforms the Dense model across all item popularity groups. In particular, the SaD framework yields substantial gains for less popular (long-tail) items. The overall improvement over conventional deep learning-based models can be largely attributed to these gains on unpopular items. For example, on the Movielens dataset, SaD achieves an approximate 25\% performance improvement for the unpopular item group, with similar trends observed on the Gowalla dataset. On the Yelp dataset, the improvement is even more pronounced. Our SaD model achieves a performance on the unpopular item group that is approximately 2.5 times higher than that of the MF model. This substantial gain not only significantly enhances recommendation accuracy for unpopular items but also effectively mitigates the impact of popularity bias.

%On the Yelp dataset, the improvement is even more pronounced. Our model achieves performance in the unpopular items group that is 2.5 times that of the MF model, significantly enhancing the recommendation performance for unpopular items and effectively mitigating the presence of popularity bias.

%The performance improvement of our model comes from two main aspects. Firstly, it enhances the performance of unpopular items. Secondly, by integrating the advantages of different views and designing the cross-view alignment module, it achieves a general performance boost across all items. By enhancing the model's ability to capture relevant patterns in less popular items, our framework ensures robust performance across a broader range of items.

The performance improvement of our model stems from two main aspects. First, it significantly enhances the recommendation accuracy for unpopular items. Second, by integrating the complementary strengths of Sparse and Dense views and employing the cross-view alignment module, it achieves a general performance boost across all items. By improving the model's ability to capture relevant patterns in less popular items, our framework ensures robust performance across a broad spectrum of items.

\subsection{Hyper-Parameter Analysis}

%In this section, we conduct a sensitivity analysis focused on the hyperparameters of our model. Through analyzing the hyperparameters, we can gain a deeper understanding of the parameters and modules within the model.

In this section, we conduct a sensitivity analysis of the key hyperparameters in our model. This analysis provides insights into how different parameter settings and module configurations affect the model’s performance, helping to better understand and fine-tune the SaD framework.

The parameter $\beta$ reflects the weighting between the Dense and Sparse models. As $\beta$ increases, the model tends more towards the Sparse model, whereas as $\beta$ decreases, it gradually leans towards the Dense model. A well-balanced consideration of the weights from the two different views leads to improved model performance. By adjusting $\beta$, we can tailor the model for different item recommendation scenarios. We observe that higher values of $\beta$ result in fewer instances of popularity bias, potentially leading to better performance for long-tail users. In practical applications, adjusting $\beta$ allows for a trade-off between performance and popularity bias.

\captionsetup[subfloat]{margin={0cm,-1cm}}

\begin{figure}[h!]
  \centering
  \subfloat[Performance on Amazon]{%
    \includegraphics[width=0.46\linewidth]{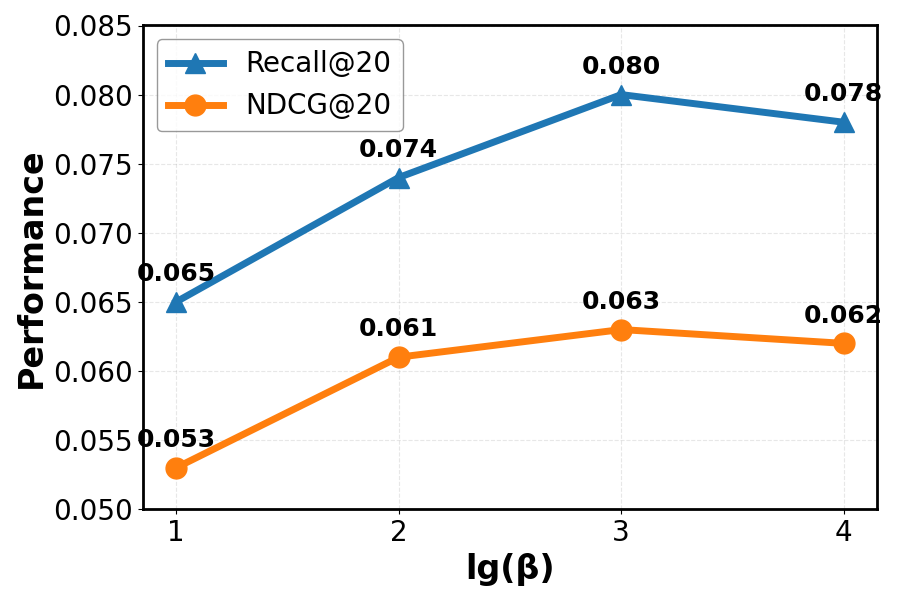}%
    \label{fig:image1}%
  }\hfill
  \subfloat[Performance on Yelp]{%
    \includegraphics[width=0.46\linewidth]{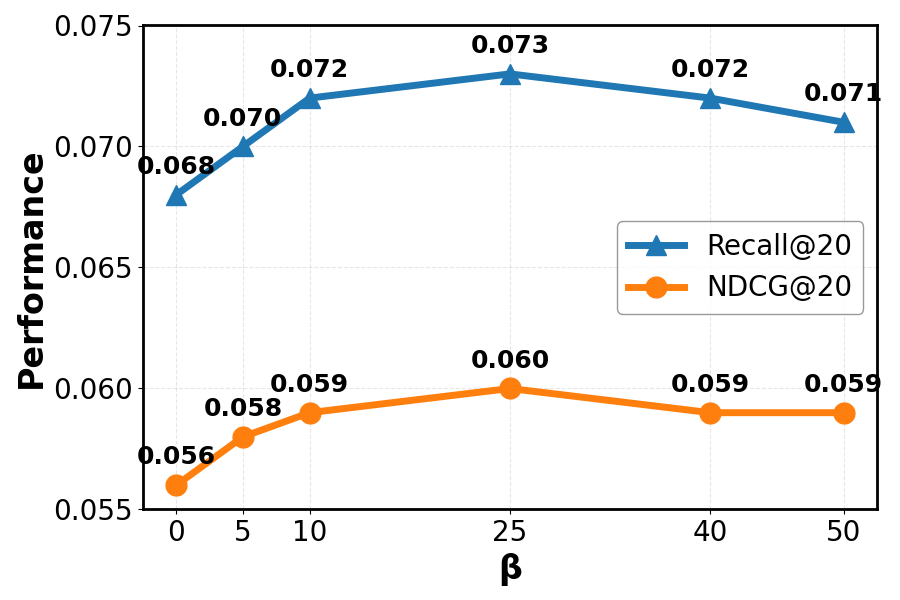}%
    \label{fig:image2}%
  } \\[0.5cm]
  \subfloat[Performance on Gowalla]{%
    \includegraphics[width=0.46\linewidth]{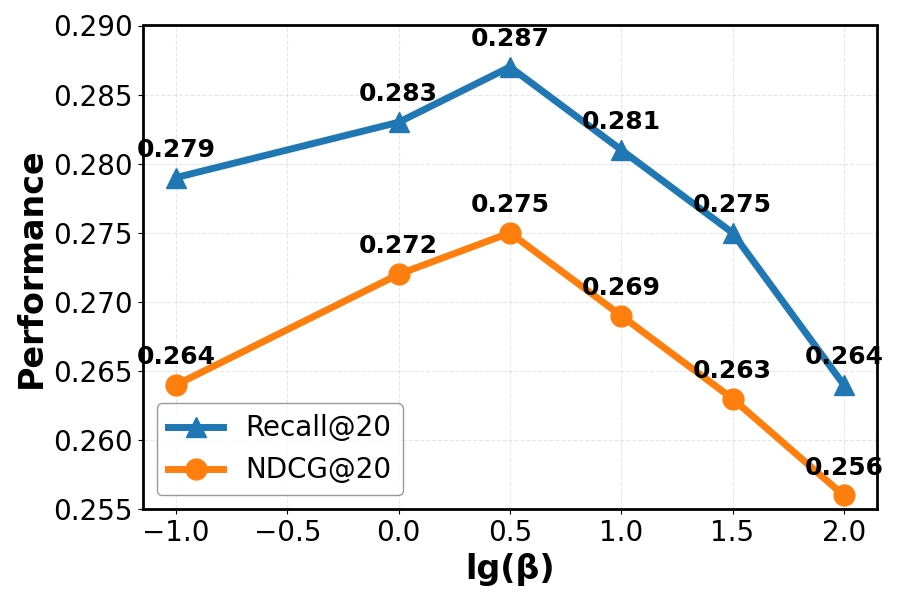}%
    \label{fig:image3}%
  }\hfill
  \subfloat[Performance on Movielens]{%
    \includegraphics[width=0.46\linewidth]{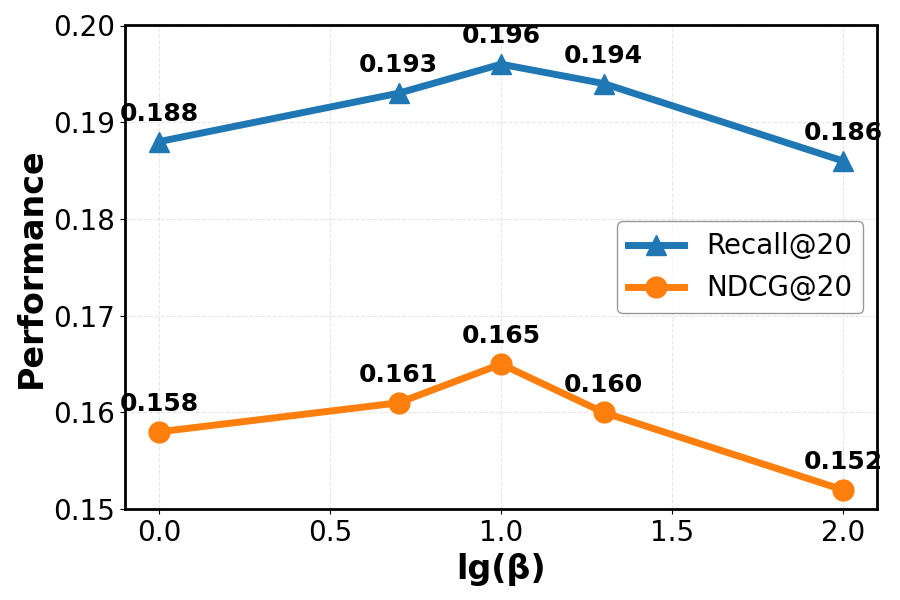}%
    \label{fig:image4}%
  }
  \caption{Performance of various $\beta$ factors.}
  \label{fig:3}
\end{figure}

Figure~\ref{fig:3} shows that on the Amazon Books and Yelp datasets, the model's performance initially improves and then declines as the $\beta$ value increases. The optimal performance is reached at $\beta = 100$ for Amazon and $\beta = 25$ for Yelp. The presence of an optimum point suggests that the $\beta$ value is a strongly correlated factor with performance. The model is quite sensitive to the value of $\beta$. Initially, increasing $\beta$ incorporates necessary explicit structural signals from the Sparse view, which effectively complements the Dense model and boosts performance. However, as $\beta$ continues to increase beyond the optimal point, the model may over-rely on the Sparse view, thereby overshadowing the semantic generalization capabilities inherent in the Dense model. This trade-off highlights that the selection of $\beta$ is critical for balancing structural robustness and semantic generalization.

To examine the robustness of the proposed dual-view alignment mechanism, we also conduct a detailed sensitivity analysis on two key hyperparameters: the proportion of pseudo-positive signals (Sparse to Dense) and the top-$K$ parameter used for neighborhood selection (Dense to Sparse).

\begin{figure}[t]
\centering
\includegraphics[width=\linewidth]{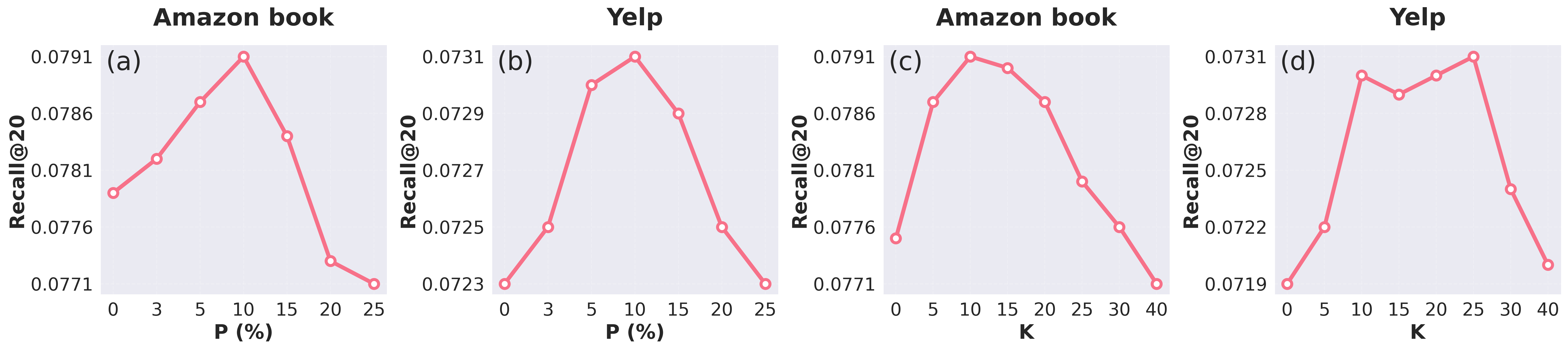}
\caption{Sensitivity analysis on Movielens and Yelp with respect to the pseudo-positive ratio and top-$K$ hyperparameters.}
\label{fig:K_sensitivity}
\end{figure}

As shown in the \emph{first two subplots} of Figure~\ref{fig:K_sensitivity}, we investigate the effect of varying the proportion of newly introduced pseudo-positive signals from 0\% to 25\%. Although the pseudo-positive rate increases, SaD consistently achieves performance gains when the proportion is within 5\%–15\%, demonstrating that the dual-view alignment effectively suppresses noise amplification and preserves useful complementary information.
The \emph{last two subplots} of Figure~\ref{fig:K_sensitivity} further analyze the influence of the top-$K$ hyperparameter. The results show that SaD maintains stable performance for $K \in [10, 30]$, and the performance degradation beyond this range remains limited, indicating that SaD achieves consistently noticeable improvements across different datasets within this range of $K$ values.
Overall, these findings confirm that SaD is robust to both the selection of $K$ and the potential pseudo-positive contamination introduced during sparse-to-dense alignment.

\subsection{Generalizability of the Model}
\label{sec:generalizability}   % 小节的唯一 label

\subsubsection{Plug-and-Play Ability}
This part demonstrates the plug-and-play capability of our model. Here, we conduct model-level experiments to demonstrate that SaD not only effectively enhances performance but is also orthogonal to the most popular self-supervised methods. In other words, our framework can be applied on top of self-supervised models to further enhance their performance.

\begin{table*}[]
  \begin{center}
  
    \caption{
SaD framework is plug-and-play: Performance boost across diverse models (Relative \% Improvement)}
    \label{tab:table5}
    \setlength{\tabcolsep}{4pt}
    \renewcommand{\arraystretch}{1.2}
    \begin{tabular}{ccccccccc}
    \toprule
      Methods & \multicolumn{2}{c}{Yelp2018} & \multicolumn{2}{c}{Gowalla} & \multicolumn{2}{c}{Amazonbooks} & \multicolumn{2}{c}{Movielens}\\
    \cline{2-3} \cline{4-5} \cline{6-7} \cline{8-9}
     & Recall@20 & NDCG@20 & Recall@20 & NDCG@20 & Recall@20 & NDCG@20 & Recall@20 & NDCG@20\\
     \midrule
    LightGCN        & 0.0605 & 0.0497 & 0.1729 & 0.1477 & 0.0385 & 0.0300 & 0.2562 & 0.2419 \\
    + SaD           & 0.0708 & 0.0584 & 0.1840 & 0.1577 & 0.0750 & 0.0595 & 0.2680 & 0.2608 \\
    $\Delta$ (\%)   & +17.0\% & +17.5\% & +6.4\% & +6.8\% & +94.8\% & +98.3\% & +4.6\% & +7.8\% \\
    \midrule
    SGL             & 0.0655 & 0.0545 & 0.1769 & 0.1502 & 0.0442 & 0.0345 & 0.2577 & 0.2507 \\
    + SaD           & 0.0721 & 0.0595 & 0.1856 & 0.1574 & 0.0756 & 0.0599 & 0.2675 & 0.2602 \\
    $\Delta$ (\%)   & +10.1\% & +9.2\% & +4.9\% & +4.8\% & +71.0\% & +73.6\% & +3.8\% & +3.8\% \\
    \midrule
    SimGCL          & 0.0724 & 0.0596 & 0.1804 & 0.1527 & 0.0444 & 0.0346 & 0.2698 & 0.2563 \\
    + SaD           & 0.0756 & 0.0622 & 0.1869 & 0.1582 & 0.0750 & 0.0597 & 0.2812 & 0.2687 \\
    $\Delta$ (\%)   & +4.4\% & +4.4\% & +3.6\% & +3.6\% & +68.5\% & +72.0\% & +4.2\% & +4.8\% \\
    \bottomrule
    
    \end{tabular}
    
  \end{center}
\end{table*}

\begin{table*}[]
\centering
\caption{Performance comparisons on additional datasets demonstrate the generalization capability of the model. Here, we use the standard benchmark dataset and report baseline results from existing works.
}\resizebox{\textwidth}{!}{
\label{tab:table6}
\renewcommand{\arraystretch}{1.2}
\begin{tabular}{ c| c c|c|cc|c| c c| c| c c}
    \toprule
    \multicolumn{3}{c}{Amazon-CDs} & \multicolumn{3}{c}{Amazon-Movies} & \multicolumn{3}{c}{Amazon-Beauty} & \multicolumn{3}{c}{Amazon-Electronics}\\
    \cline{1-3} \cline{4-6} \cline{7-9} \cline{10-12}
     Method &Recall & NDCG &Method & Recall & NDCG & Method &Recall & NDCG & Method &F1 & NDCG\\
    \midrule
    NGCF & 0.1258 &0.0792  &NGCF & 0.0866 & 0.0555& MF-BPR & 0.1312&0.0778&MF-BPR&0.0275&0.0680 \\
    NIA-GCN  & 0.1487 &0.0932 &  NIA-GCN & 0.1058 & 0.0683 & NIA-GCN & 0.1513 & 0.0917&ENMF&0.0314&0.0823\\
    BGCF & 0.1506 &0.0948 & BGCF & 0.1066 & 0.0693 & BGCF & 0.1534 & 0.0912&NBPO&0.0313&0.0810\\
    SimpleX & 0.1763 &0.1145 & SimpleX & 0.1342 & 0.0926 & SimpleX &0.1721& 0.1028&UltraGCN&0.0330&0.0829\\
    slim & 0.1605 & 0.1098  & slim &0.1288  & 0.0920 & slim &  0.1667 & 0.1083&slim&0.0260&0.0646\\
    \midrule
    \rowcolor{gray!15} SaD       & \textbf{0.1806}        & \textbf{0.1194}           &SaD        & \textbf{0.1447} & \textbf{0.1019}& SaD & \textbf{0.1802} & \textbf{0.1135}  &SaD&\textbf{0.0343} &\textbf{0.0843}   \\
    \bottomrule
  \end{tabular}}
\end{table*}

Since existing methods are largely based on Dense models, we evaluate the generalizability of our framework by replacing the Dense component with several classic model architectures and measuring performance.

%We select LightGCN, along with SGL and SimGCL, which incorporate self-supervised learning into LightGCN, as baselines. We test the performance of our model when these architectures were used as the Dense model. We pass the model as Dense into the framework and train the SaD model using the same steps.

Specifically, we select LightGCN, along with SGL and SimGCL, which incorporate self-supervised learning into LightGCN, as alternative Dense models. We integrate each of these architectures into the SaD framework and train the model following the same procedure, enabling a consistent comparison across different backbones.

%To ensure the generalizability of our model, we simply select Matrix Factorization (MF) as the Dense model for the primary experiments. However, our model can also be applied to Dense models based on graphs or other methods. As shown in Table~\ref{tab:table5}, all Dense models exhibited significant performance improvements within our framework. The data indicate that, unlike the performance gains achieved through self-supervised loss, the improvements brought by our model are orthogonal to self-supervision. Our model can be applied to a wider range of other models.

To demonstrate the generalizability of our framework, we use Matrix Factorization (MF) as the Dense model in the primary experiments. Importantly, the SaD framework can also accommodate Dense models based on graphs or other architectures. As shown in Table~\ref{tab:table5}, integrating various Dense models into our framework consistently yields substantial performance improvements. These results indicate that, unlike gains obtained from self-supervised loss, the improvements introduced by our model are largely orthogonal to self-supervision. Overall, the SaD framework is flexible and can enhance a wide range of existing recommendation models.

%Our framework significantly enhances model performance. On LightGCN, our model effectively improves performance. Additionally, comparisons with methods like SGL and SimGCL show that the performance gains from our model are independent of those from self-supervised methods. On the Yelp dataset, even the state-of-the-art method SimGCL achieves an approximate 5\% performance improvement when integrated with our framework. For simplicity, we only use MF as the dense method in our paper. Using other more complex dense models would likely yield even better performance.

Our framework substantially enhances model performance across different Dense architectures. When applied to LightGCN, the SaD framework effectively improves recommendation quality. Furthermore, comparisons with self-supervised methods such as SGL and SimGCL demonstrate that the performance gains provided by our model are largely orthogonal to those achieved through self-supervision. For instance, on the Yelp dataset, even the state-of-the-art SimGCL achieves an additional $\sim$5\% improvement when integrated within our framework. For clarity and simplicity, we primarily use MF as the Dense model in this paper; employing more advanced Dense architectures would likely result in even greater performance gains.

\begin{figure}[t]
  \centering
  \includegraphics[width=\linewidth]{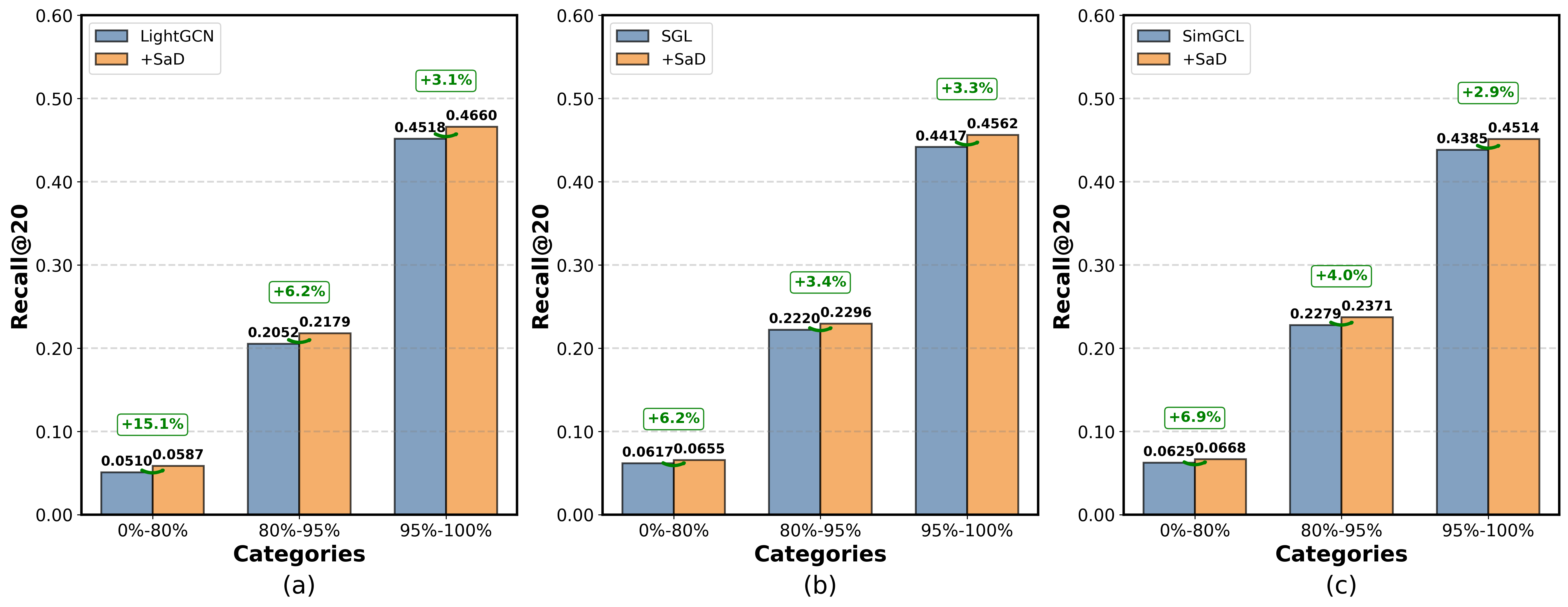}
  \caption{Recall@20 across degree groups (Low/Mid/High) for three backbones, with and without SaD.}
  \label{fig:gnn_extension}
\end{figure}

Additionally, we conduct a fine-grained mechanism-level analysis to further confirm the generalizability of SaD. As shown in Figure~\ref{fig:gnn_extension}, SaD consistently improves performance across user groups with different interaction degrees, indicating that the benefit is not tied to any specific backbone or user distribution pattern. These results further validate the robustness of SaD and demonstrate its potential to serve as a general enhancement framework for various recommendation architectures.

\subsubsection{Evaluating SaD on Additional Datasets}

%To demonstrate the generalization ability of our SaD model, we extend our model's performance evaluation to include a wider array of datasets. Testing its performance on a broader range of datasets allows for a better demonstration of the contribution and practical value of our model.
To further demonstrate the generalization capability of our SaD model, we evaluate its performance on a broader set of datasets. Assessing the model across diverse data enables a more comprehensive understanding of its contributions and practical applicability.

In Table~\ref{tab:table6}, we show the performance of several state-of-the-art models on diverse datasets chosen to evaluate generalization capabilities beyond classic benchmarks. We maintain consistent dataset splits and evaluation metrics as in previous studies~\cite{mao2021ultragcn,mao2021simplex}.

%This consistent performance across a diverse range of datasets highlights its strong generalization capability, indicating that the model can effectively adapt to different types of data and tasks. With its ability to maintain competitive results on all eight datasets, our model proves to be both reliable and scalable, offering significant advantages in real-world applications where data can vary widely.

This consistent performance across a diverse set of datasets underscores the strong generalization capability of our model, demonstrating its ability to effectively adapt to different types of data and tasks. By maintaining competitive results across all eight datasets, our model proves to be both reliable and scalable, offering notable advantages in real-world applications where data characteristics can vary widely.

\subsection{Empirical Validation for Theoretical Insights}

%To empirically validate the theoretical insights, we conduct detailed experiments to examine the consistency between the theoretical predictions and observed model behavior. For each user $u$ and test positive $i^+$, we sample $K$ negatives $i^-$ uniformly from unobserved items (excluding all train/test positives). Bucket SNR is computed by \emph{conditioning on the positive’s bucket}; negatives are drawn from the full candidate pool. This protocol evaluates stability under realistic candidate hardness.

To empirically validate our theoretical findings, we conduct detailed experiments to examine the consistency between predicted and observed model behaviors. For each user $u$ and test positive item $i^+$, we sample $K$ negative items $i^-$ uniformly from unobserved items (excluding all train and test positives). The bucketed SNR is computed by \emph{conditioning on the positive item's bucket}, while negatives are drawn from the full candidate set. This protocol evaluates the model's stability under realistic candidate difficulty.

We perform the experiments on the MovieLens-1M dataset, and the results are summarized as follows: (i) \textbf{Sparse has the cleanest raw signal on tail:} SNR $2.45$ vs.\ $1.79$ (dense).  
(ii) \textbf{Alignment improves the dense branch on tail and the overall Recall:} SaD raises tail SNR to $2.05$ for the dense branch and overall Recall@20 to $0.2865$ (dense $0.2777$, sparse $0.2577$). While sparse-alone still has the highest tail SNR, dual alignment offers a better Pareto trade-off between stability and recommendation quality.  
(iii) \textbf{Complementarity increases on head:} On head items, SaD achieves a higher SNR ($1.0378$) compared to dense ($0.9341$) and sparse ($1.0364$) models, indicating that the dual-view fusion effectively enhances complementarity even in popular-item regions.
(iv) \textbf{Alignment enhances the overall SNR even with increased correlation~($\rho$):} Without alignment (SaD w/o align), the correlation between the sparse and dense views is lower, but this uncoordinated diversity leads to weaker overall SNR and reduced recall. 
Although bidirectional alignment slightly increases the inter-view correlation~($\rho$), it substantially boosts each view’s individual SNR, resulting in a significant overall SNR improvement and better retrieval performance.

\begin{table}[ht]
  \centering
  \caption{
  Quantitative validation on MovieLens-1M. 
  We report SNR (higher is better) for three item-popularity buckets—long-tail (0--80\%), mid (80--95\%), head (95--100\%)—and overall results. 
  For fusion models, we also show the cross-view Pearson correlation~$\rho$.
  }
  \setlength{\tabcolsep}{6pt}
  \renewcommand{\arraystretch}{1.2}
  \resizebox{\linewidth}{!}{%
    \begin{tabular}{lcccccc}
      \toprule
      \textbf{Model} &
      \textbf{SNR (Long-tail)} &
      \textbf{SNR (Mid)} &
      \textbf{SNR (Head)} &
      \textbf{SNR (Overall)} &
      \textbf{Recall@20 (Overall)} &
      \textbf{Corr ($\rho$)} \\
      \midrule
      Dense & 1.7912 & 1.3602 & 0.9341 & 1.6837 & 0.2777 & N/A \\
      Sparse & 2.4489 & 1.4892 & 1.0364 & 2.2343 & 0.2577 & N/A \\
      SaD w/o align & 1.9232 & 1.4209 & 0.9821 & 1.8008 & 0.2847 & 0.24 \\
      \textbf{SaD} & \textbf{2.0466} & \textbf{1.5015} & \textbf{1.0378} & \textbf{1.9144} & \textbf{0.2865} & \textbf{0.26} \\
      \bottomrule
    \end{tabular}%
  }
  \label{tab:theory_comparison}
\end{table}

%% file: Ours/chapter/conclusion.tex
%In our study, we introduce the SaD model, a novel approach that combines Dual Views to enhance collaborative filtering tasks. This integration promotes effective information exchange between the two Views, resulting in superior model performance. The model can significantly improve the performance of various dense models. The SaD model excels in accuracy, as demonstrated across a variety of datasets, indicating its strong potential for broad application in different settings. Its effectiveness across various datasets suggests potential for broad generalization. In the future, more combination strategies can be explored. Enhancements might include optimizing the information exchange processes or incorporating advanced machine learning techniques to better capture the complexities of user-item interactions. We will also consider the use of a learned gate for combining the two views and plan to investigate this adaptive fusion mechanism in future work. Meanwhile, the model can be extended to social recommendation, multi-behavior recommendation and other recommendation areas.

In this study, we introduce the SaD model, a novel framework that integrates Dual Views to enhance collaborative filtering. By facilitating effective information exchange between the Sparse and Dense views, the model achieves superior performance across a range of dense model backbones. Extensive experiments on multiple datasets demonstrate the SaD model's accuracy and robustness, highlighting its strong potential for broad generalization. 

Future work may explore additional combination strategies, such as optimizing the information exchange mechanisms or incorporating advanced machine learning techniques to better capture complex user-item interactions. We also plan to investigate adaptive fusion mechanisms, including learned gates for dynamically combining the two views. Furthermore, the SaD framework can be extended to other recommendation settings, such as social recommendation and multi-behavior recommendation, demonstrating its versatility and applicability.

\section*{Acknowledgments}
This research was supported by the Public Computing Cloud of Renmin University of China and by the Fund for Building World-Class Universities (Disciplines) at Renmin University of China.